\documentstyle[11pt]{article}
\begin{document}
\newcommand{\s}{\varsigma}
\newcommand{\room}{\rule[-0.3cm]{0cm}{0.8cm}}
\newcommand{\hsp}{\hspace*{3mm}}
\newcommand{\vsp}{\vspace*{3mm}}
\newcommand{\be}{\begin{equation}}
\newcommand{\ee}{\end{equation}}
\newcommand{\bd}{\begin{displaymath}}
\newcommand{\ed}{\end{displaymath}}
\newcommand{\bdm}{\begin{displaymath}}
\newcommand{\edm}{\end{displaymath}}
\newcommand{\bea}{\begin{eqnarray}}
\newcommand{\eea}{\end{eqnarray}}
\newcommand{\sgn}{~{\rm sgn}}
\newcommand{\extr}{~{\rm extr}}
\newcommand{\Equiv}{\Longleftrightarrow}
\newcommand{\pprime}{\prime\prime}
\newcommand{\notexists}{\exists\hspace*{-2mm}/}
\newcommand{\bra}{\langle}
\newcommand{\ket}{\rangle}
\newcommand{\bigbra}{\left\langle\room}
\newcommand{\bigket}{\right\rangle\room}
\newcommand{\bras}{\langle\!\langle}
\newcommand{\kets}{\rangle\!\rangle_{xy}}
\newcommand{\bigbras}{\left\langle\!\!\!\left\langle\room}
\newcommand{\bigkets}{\right\rangle\!\!\!\right\rangle_{\!\!xy}\room}
\newcommand{\order}{{\cal O}}
\newcommand{\minus}{\!-\!}
\newcommand{\plus}{\!+\!}
\newcommand{\erf}{{\rm erf}}
\newcommand{\bk}{\mbox{\boldmath $k$}}
\newcommand{\bm}{\mbox{\boldmath $m$}}
\newcommand{\br}{\mbox{\boldmath $r$}}
\newcommand{\bq}{\mbox{\boldmath $q$}}
\newcommand{\bz}{\mbox{\boldmath $z$}}
\newcommand{\bH}{\mbox{\boldmath $H$}}
\newcommand{\bM}{\mbox{\boldmath $M$}}
\newcommand{\bQ}{\mbox{\boldmath $Q$}}
\newcommand{\bR}{\mbox{\boldmath $R$}}
\newcommand{\bW}{\mbox{\boldmath $W$}}
\newcommand{\hbh}{\hat{\mbox{\boldmath $h$}}}
\newcommand{\hbm}{\hat{\mbox{\boldmath $m$}}}
\newcommand{\hbr}{\hat{\mbox{\boldmath $r$}}}
\newcommand{\hbq}{\hat{\mbox{\boldmath $q$}}}
\newcommand{\hbD}{\hat{\mbox{\boldmath $D$}}}
\newcommand{\hbQ}{\hat{\mbox{\boldmath $Q$}}}
\newcommand{\hbR}{\hat{\mbox{\boldmath $R$}}}
\newcommand{\hbW}{\hat{\mbox{\boldmath $W$}}}
\newcommand{\bsigma}{\mbox{\boldmath $\sigma$}}
\newcommand{\bomega}{\mbox{\boldmath $\Omega$}}
\newcommand{\bphi}{\mbox{\boldmath $\Phi$}}
\newcommand{\bpsi}{\mbox{\boldmath $\psi$}}
\newcommand{\bdelta}{\mbox{\boldmath $\Delta$}}
\newcommand{\btheta}{\mbox{\boldmath $\theta$}}
\newcommand{\bxi}{\mbox{\boldmath $\xi$}}
\newcommand{\bmu}{\mbox{\boldmath $\mu$}}
\newcommand{\brho}{\mbox{\boldmath $\rho$}}
\newcommand{\bEta}{\mbox{\boldmath $\eta$}}
\newcommand{\req}{r_{\rm eq}}
\newcommand{\unity}{{\bf 1}\hspace{-1mm}{\bf I}}

\title{\bf Order-Parameter Flow in the SK Spin-Glass\\ II: Inclusion of
Microscopic Memory Effects}
\author{S.N. Laughton \and A.C.C. Coolen \dag \and D. Sherrington}
\maketitle

\begin{center}
Department of Physics - Theoretical Physics\\
University of Oxford\\
1 Keble Road, Oxford OX1 3NP, U.K.
\end{center}
\vsp

\begin{center} PACS: 75.10N, 05.20 \end{center}\vsp
\vsp

\begin{abstract}\noindent
We develop further a recent dynamical replica theory to describe
the dynamics of the Sherrington-Kirkpatrick spin-glass in terms of closed
evolution equations for macroscopic order parameters. We show how
microscopic memory effects can be included in the formalism through
the introduction  of a dynamic order parameter function: the joint
spin-field distribution.
The resulting formalism describes very accurately the relaxation
phenomena observed in numerical simulations,
including the typical overall slowing down of the
flow that was missed by the previous simple
two-parameter theory.
The advanced dynamical replica
theory is either exact or a very good approximation.
\end{abstract}
\vspace*{35mm}

\noindent
\dag~ present address:\\
Dept. of Mathematics, King's College London\\
Strand, London WC2R 2LS, U.K.

\pagebreak\tableofcontents

\pagebreak\section{Introduction}

In a previous paper \cite{sk1} we introduced a theory to describe
the Glauber dynamics of the Sherrington-Kirkpatrick (SK) \cite{SK}
spin-glass model in terms of deterministic flow equations for two
macroscopic state variables: the
magnetisation $m$ and the spin-glass contribution $r$ to the
energy (for a more general discussion of the SK model
and of the relevant literature we refer to \cite{sk1}).
The theory is
based on the removal of microscopic memory
effects: the only `knowledge' the system is assumed to have of its
past is the value of the macroscopic state $(m,r)$.
In fact any macroscopic
dynamical theory for the SK model must contain as dynamical variables,
either explicitly or implicitly,
at least the magnetisation $m$ (which is the relevant observable for
many typical
spin-glass remanence phenomena) and the total energy of the system (in
order for the theory to reproduce the correct equilibrium
equations). Since for
the SK model the energy per spin is a simple function of the
macroscopic state vector $(m,r)$, the theory of \cite{sk1} can be seen as the
simplest two-parameter dynamical theory for the SK model that has the
properties of being exact for short times (upon choosing appropriate
initial conditions) and in equilibrium.

At a technical level, the resulting formalism is a dynamical replica
theory, which at fixed-points of the macroscopic flow reduces to the
standard equilibrium replica theory, including replica symmetry
breaking (RSB) \`{a} la Parisi \cite{parisi} if it occurs. This is in
contrast to an alternative formalism based on path-integral
methods (see e.g. \cite{zinnjustin,sommers}), where it is not yet
known
how to recover the standard
equilibrium results in RSB situations. In fact, the potential of the
present theory to provide the link between equilibrium replica theory
and the description in terms of correlation- and response functions
(once the hitherto neglected microscopic memory effects have been
incorporated), we regard as one of its most interesting features.

In \cite{sk1} the actual macroscopic flow equations were calculated
only within the replica-symmetry (RS) ansatz. The RS version of
the theory was quite succesful at predicting the flow trajectories in
the $(m,r)$ plane, but also exhibited clear deviations in terms of the
long-term temporal dependence of the macroscopic state variables $m$
and $r$. These are partly
due to the elimination from the dynamical equations  of microscopic
memory effects, and partly an artefact of the RS ansatz. The latter
cause for deviations can be dealt with by allowing for a breaking of
replica symmetry, following Parisi's RSB scheme
\cite{parisi}; although technically non-trivial,
this can be seen as a straightforward generalisation of the theory
(dynamical RSB within the present formalism will be the subject of a
subsequent paper \cite{sk3}). Here
we
concentrate on the more subtle question of how to incorporate
microscopic memory
effects, i.e. on how to generalise the ideas in \cite{sk1} to a
situation where a macroscopic state specifies the details of the
underlying microscopic  states
to a much higher degree.
We will show that, by considering as the appropriate macroscopic dynamical
observable
the joint spin-field distribution, one can indeed follow the
steps in \cite{sk1} and arrive at a dynamic replica theory
which not only inherits
by construction the exactness of the previous
(simple) two-parameter theory in the temporal limits $t\rightarrow0$
and $t\rightarrow\infty$, but also describes the simulation
experiments accurately, as far as our (limited) data allow us to
conclude. The philosophy of our approach resembles the one proposed
by Horner \cite{hornerfielddist}, who also derived a closed diffusion
equation for the evolution of the
joint spin-field distribution. However, at the technical level of the
closure procedure there are
important differences.  The present theory is not only exact
for short times (as is \cite{hornerfielddist}), but also in
equilibrium, in the sense that the {\em full} RSB equations are
recovered. More importantly, it is constructed in such a way that it will
produce exact dynamic equations if the joint spin-field distribution indeed
turns out to constitute a closed level of description.

This paper is organised as follows. First we generalise the theory of
\cite{sk1} to the case of an arbitrary set of macroscopic observables,
and derive constraints on the allowed choices for such a set, by
requiring exactness in specific limits. We then derive a diffusion
equation for the joint spin-field distribution, which generates a
dynamical replica theory. The relevant equations are simplified by
making the RS ansatz, and their predictions are compared to the
results of numerical simulations. We close the paper with a discussion
of our results and their implications.

\section{Dynamics \& Replicas}

\subsection{Definitions and Macroscopic Laws}

The Sherrington-Kirkpatrick (SK) spin-glass model \cite{SK} consists
of  $N$ Ising spins
$\sigma_{i}\in\{-1,1\}$ with infinite-range exchange
interactions $J_{ij}$:
\be
J_{ij}=\frac{1}{N}J_0+\frac{1}{\sqrt{N}}J
z_{ij}~~~~~~~~(i<j)
\label{eq:interactions}
\ee
where the quantities $z_{ij}$, which represent quenched disorder,  are
drawn independently at random
from a Gaussian distribution with zero mean and unit variance.
The evolution in time of the microscopic probability distribution
$p_{t}(\bsigma)$ is taken to be of the Glauber form described by
the master equation
\be
\frac{d}{dt}p_{t}(\bsigma)=\sum_{k=1}^{N}
\left[p_{t}(F_k\bsigma)w_{k}(F_k\bsigma)-p_{t}(\bsigma)w_{k}(\bsigma)\right]
\label{eq:master}
\ee
in which $F_k$ is a spin-flip operator
$F_{k}\Phi(\bsigma)\equiv\Phi(\sigma_1,\ldots,-\sigma_k,\ldots,\sigma_N)$ and
the transition rates $w_k(\bsigma)$ and the local alignment fields
$h_i(\bsigma)$ are
\be
w_{k}(\vec{s})=\frac{1}{2}\left[1-\sigma_k\tanh[\beta
h_k(\bsigma)]\right]~~~~~~~~
h_{i}(\bsigma)=\sum_{j\neq i}J_{ij}\sigma_{j}+\theta
\label{eq:ratesandfields}
\ee
where $\beta$ is the inverse temperature,
which leads asymptotically to the required standard Boltzmann equilibrium
distribution $p_\infty(\bsigma)\sim \exp[\minus\beta H(\bsigma)]$,
with the conventional Hamiltonian
\be
H(\bsigma)= -\sum_{i< j}\sigma_i J_{ij}\sigma_j-\theta\sum_i\sigma_i
\label{eq:hamiltonian}
\ee

We now turn to the evolution  in time of any given set of $\ell$ macroscopic
observables $\bomega (\bsigma)=
( \Omega_1(\bsigma), \dots, \Omega_\ell(\bsigma))$,
described by the macroscopic probability
distribution
${\cal P}_t (\bomega) = \sum_{\bsigma} p_t (\bsigma) \delta\left[
\bomega - \bomega (\bsigma)\right]$.
We insert the master equation
(\ref{eq:master}), and expand the result in powers of the
`discrete derivatives' $\Delta^\mu_i (\bsigma) =
\Omega_\mu (F_i \bsigma)\minus \Omega_\mu (\bsigma)$, which gives
\bd
	\frac{d }{d t}  {\cal P}_t (\bomega) =
- \sum_{\mu=1}^{\ell}
\frac{\partial }{\partial \Omega_\mu} \left\{ {\cal P}_t (\bomega)
\bra\sum_i w_i (\bsigma) \Delta^\mu_i (\bsigma)
\ket_{\bomega,t} \right\}
\ed
\be
+\frac{1}{2}\sum_{\mu\nu=1}^{\ell}
\frac{\partial^2 }{\partial \Omega_\mu\partial\Omega_\nu}
\left\{ {\cal P}_t (\bomega)
\bra\sum_i w_i (\bsigma) \Delta^\mu_i (\bsigma)\Delta^\nu_i (\bsigma)
\ket_{\bomega,t} \right\}
+ {\cal O} (N \ell^3 \Delta^3)
\label{eq:macromaster}
\ee
where we introduced the sub-shell, or conditional, average
\bd
\bra f (\bsigma) \ket_{\bomega,t} =
\frac{\sum_{\bsigma} p_t (\bsigma) \, \delta\left[\bomega \minus \bomega
(\bsigma)\right] \, f (\bsigma)}
{\sum_{\bsigma} p_t (\bsigma) \, \delta\left[\bomega \minus \bomega
(\bsigma)\right]}
\ed
If the second (diffusion) term, which is ${\cal O} (N \ell^2
\Delta^2)$,
vanishes for
$N\rightarrow\infty$, equation (\ref{eq:macromaster}) acquires the
Liouville form, the solution of which describes the deterministic flow
\be
	\frac{d }{d t} \bomega_t = \bra
\sum_i w_i (\bsigma)\left[
\bomega(F_i\bsigma)\minus \bomega(\bsigma) \right]\ket_{\bomega_t,t}
\label{eq:omegaflow}
\ee
Although exact for $N\rightarrow\infty$ (provided the diffusion term
indeed vanishes), the set (\ref{eq:omegaflow}) need not be closed, due to the
appearance of $p_t(\bsigma)$ in the
sub-shell average.

There are two {\em natural} ways for the set (\ref{eq:omegaflow}) to
close.
Firstly, by the argument of the subshell average in
(\ref{eq:omegaflow}) depending on
$\bsigma$ only through $\bomega(\bsigma)$ (now $p_t(\bsigma)$ will
simply drop out), and secondly by the microscopic dynamics
(\ref{eq:master}) allowing for equipartitioning solutions (where
$p_t(\bsigma)$ depends on $\bsigma$ only through $\bomega(\bsigma)$).
In both cases one obtains the correct equations for $\bomega_t$
upon simply eliminating $p_t(\bsigma)$ from (\ref{eq:omegaflow}).

\subsection{Closed Flow Equation for an Order Parameter Function}

Generalising
\cite{sk1} to the present case, we now make the
following assumptions:
\begin{enumerate}
\item The observables $\bomega(\bsigma)$ are self-averaging with respect to
the microscopic realisation of the disorder $\{z_{ij}\}$, at any time.
\item In evaluating the sub-shell averages we assume equipartitioning
of probability within the $\bomega$-subshell of the ensemble.
\end{enumerate}
As a result $p_t(\bsigma)$ drops out, and the macroscopic
equations (\ref{eq:omegaflow}) are replaced
by closed ones, from which the unpleasant fraction is removed as
in \cite{sk1} using the
replica identity
\be
\frac{\sum_{\bsigma} \Phi ({\bsigma}) W
({\bsigma})}{\sum_{\bsigma} W ({\bsigma})} = \lim_{n \rightarrow 0}
\sum_{{\bsigma}^1} \cdots \sum_{{\bsigma}^n} \Phi ({\bsigma}^1)
\prod_{\alpha=1}^n W ({\bsigma}^\alpha)
\label{eq:replicaidentity}
\ee
We now obtain:
\bd
\frac{d}{dt} \bomega_t =
\bra
\frac{\sum_{\bsigma}\delta\left[\bomega\minus\bomega(\bsigma)\right]
\sum_i w_i (\bsigma)\left[\bomega(F_i\bsigma)\minus \bomega(\bsigma) \right]}
{\sum_{\bsigma}\delta\left[\bomega\minus\bomega(\bsigma)\right]}
\ket_{\{z_{ij}\}}~~~~~~~~~~~~~~~~~~~~
\ed
\be
{}~~~~~~=
\lim_{n\rightarrow0} \sum_{\bsigma^1}\cdots\sum_{\bsigma^n}
\bra
\sum_i w_i (\bsigma)
\left[\bomega(F_i\bsigma^1)\minus \bomega(\bsigma^1) \right]
\prod_{\alpha=1}^n
\delta\left[\bomega\minus\bomega(\bsigma^\alpha)\right]
\ket_{\{z_{ij}\}}
\label{eq:closedomegaflow}
\ee
Our second assumption (equipartitioning) is the dangerous one;
its impact on the
accurateness of the theory depends critically on the choice made
for the observables $\bomega(\bsigma)$. If, however, the
observables $\bomega(\bsigma)$ indeed obey closed self-averaging
dynamic equations, our closure procedure will be exact
(see our reasoning above).
Requiring the theory to be
exact in two solvable limits (equilibrium and
$J\rightarrow0$, respectively) imposes
constraints
on the allowed choices for $\bomega(\bsigma)$. Since in equilibrium we have
equipartitioning of probability in the energy-subshells (with the
Hamiltonian (\ref{eq:hamiltonian})), and since for
$J\rightarrow 0$ one obtains closed dynamic equations for the
magnetisation, our two requirements imply
$\bomega(\bsigma)=(m(\bsigma),H(\bsigma),\ldots)$ (modulo equivalent
combinations).
For the SK model the energy per spin can be written as
\be
H(\bsigma)/N= -\frac{1}{2}J_0 m^2(\bsigma)-\theta m(\bsigma)
-Jr(\bsigma)+\frac{1}{2}J_0/N
\label{eq:energyperspin}
\ee
with
\be
m(\bsigma)=\frac{1}{N}\sum_i\sigma_i
{}~~~~~~~~r(\bsigma)=\frac{1}{N\sqrt{N}}\sum_{i<j}\sigma_i
z_{ij}\sigma_j
\label{eq:orderparameters}
\ee
so the choice made in \cite{sk1} leads to the simplest
two-parameter theory that meets our requirements of exactness in the
two solvable limits. Improving upon \cite{sk1} implies including
microscopic information beyond $(m,r)$, i.e. adding observables to the
set $\bomega(\bsigma)=(m(\bsigma),r(\bsigma))$. Addition of any finite
number of observables, although technically simple, is not
expected to give more than just minor corrections. In contrast we
choose for the set of observables $\bomega(\bsigma)$ the (infinite
dimensional) joint
spin-field distribution:
\be
D(\s,h;\bsigma)
= \frac{1}{N} \sum_i \delta_{\s,\sigma_i}
\delta\left[h \minus h_i (\bsigma)\right]
\label{eq:distribution}
\ee
with the local fields (\ref{eq:ratesandfields}).
Our motivation for this choice is the following
\begin{enumerate}
\item The previous two dynamic parameters $m(\bsigma)$ and
$r(\bsigma)$
can be written
as integrals over $D(\s,h;\bsigma)$, so the advanced theory
automatically inherits the
exactness in the two solvable limits $t\rightarrow\infty$ and
$J\rightarrow0$.
\item The order parameter function $D(\s,h)$ specifies the
underlying states $\bsigma$ to a much higher degree than
$(m,r)$; i.e. microscopic memory is taken into account.
\item The microscopic equation (\ref{eq:master}) itself is formulated
in terms of spins and fields.
\item The choice (\ref{eq:distribution}) allows for immediate
generalisation to models without detailed balance and to soft-spin models.
\end{enumerate}
To avoid all kinds of technical difficulties we assume that the distribution
(\ref{eq:distribution}) is sufficiently well behaved; we assume that
we can evaluate $D_t(\s,h)$ in a number $\ell$ of field arguments
$h_\mu$ and take the limit
$\ell\rightarrow\infty$ {\em after} the limit
$N\rightarrow\infty$. We then have $2\ell$ observables
$\Omega_{\s\mu}(\bsigma)=D(\s,h_\mu;\bsigma)$, with
$\mu=1,\ldots,\ell$ and $\s=\pm 1$.
In order to work out equation (\ref{eq:closedomegaflow}) we calculate the
discrete derivatives $\Delta^{\s\mu}_i (\bsigma)
=D(\s,h_\mu;F_i\bsigma)
\minus  D(\s,h_\mu;\bsigma)$:
\bd
\Delta^{\s\mu}_i (\bsigma) = \frac{2\sigma_i}{N\sqrt{N}}
\sum_{j\neq i}\delta_{\s,\sigma_j}\delta^{\prime}[h_\mu\minus h_j(\bsigma)]
\left[\frac{J_0}{\sqrt{N}}\plus Jz_{ij}\right]
\ed
\be
+\frac{2J^2}{N^2}
\sum_{j\neq i}\delta_{\s,\sigma_j}\delta^{\pprime}[h_\mu\minus
h_j(\bsigma)]z_{ij}^2
-\frac{1}{N}\s \sigma_i\delta[h_\mu\minus
h_i(\bsigma)]+\order(N^{-\frac{3}{2}})
\label{eq:delta}
\ee
where primes indicate derivatives (in a distributional sense).
Since $\Delta^{\s\mu}_i(\bsigma)=\order(N^{-\frac{1}{2}})$,
the diffusion term in
(\ref{eq:macromaster}) could be $\order(1)$. Explicit calculation,
however, will show that it vanishes as $N^{-\frac{1}{2}}$; at this
stage we anticipate that calculation and assume deterministic
evolution. To suppress notation we write
\bd
\sum_{\sigma}\int\!dH~f(\sigma,H) D(\sigma,H)=\bra f(\sigma,H)\ket_D
\ed
In order to see clearly for which terms our two closure
assumptions will actually be operational, we first work out the exact
equation (\ref{eq:omegaflow}). We insert (\ref{eq:delta}) into
(\ref{eq:omegaflow}) and retain only the leading $\order(1)$
terms:
\bd
\frac{\partial}{\partial t} D_t(\s,h) =
\frac{1}{2}\left[1\plus\s\tanh(\beta h)\right]D_t(\minus\s,h)
-\frac{1}{2}\left[1\minus\s\tanh(\beta h)\right]D_t(\s,h)
\ed
\bd
+\frac{\partial}{\partial h} \left\{
D_t(\s,h) [h\minus\theta\minus J_0\bra \tanh(\beta H)\ket_{D_t}]
- \frac{J}{N\sqrt{N}} \sum_{i\neq j} z_{ij}
\bra\tanh(\beta h_i(\bsigma)) \delta_{\s,\sigma_j}
\delta[h\minus h_j(\bsigma)] \ket_{D_t;t}
\right\}
\ed
\be
 + J^2 \frac{\partial^2 }{\partial h^2} \left\{
\frac{1}{N^2}\sum_{i\neq j}z^2_{ij}\bra\left[1\minus\sigma_i\tanh(\beta
h_i(\bsigma)) \right]
\delta_{\s,\sigma_j} \delta[h \minus h_j (\bsigma)]
\ket_{D_t; t} \right\}
+ {\cal O} (N^{-\frac{1}{2}})
\label{eq:exactdiffusion}
\ee
with the sub-shell average
\be
\bra f (\bsigma)\ket_{D;t}=
\frac{
\sum_{\bsigma} p_t(\bsigma) f (\bsigma)
\prod_{\s\mu}
\delta\left[D(\s,h_\mu)\minus
\frac{1}{N}\sum_j \delta_{\s,\sigma_j}
\delta\left[h_\mu\minus h_j (\bsigma) \right]\right]}
{\sum_{\bsigma} p_t(\bsigma)
\prod_{\s\mu}\delta \left[D(\s,h_\mu)\minus
\frac{1}{N}\sum_j \delta_{\s,\sigma_j}
\delta\left[h_\mu\minus h_j (\bsigma) \right]\right]}
\label{eq:subshellaverage}
\ee
The {\em closed} dynamic equation (\ref{eq:closedomegaflow}) is subsequently
obtained from (\ref{eq:exactdiffusion}) by elimination of
$p_t(\bsigma)$ and averaging over the disorder (using identity
(\ref{eq:replicaidentity})):
\bd
\frac{\partial}{\partial t} D_t(\s,h) =
\frac{1}{2}\left[1\plus\s\tanh(\beta h)\right]D_t(\minus\s,h)
-\frac{1}{2}\left[1\minus\s\tanh(\beta h)\right]D_t(\s,h)
\ed
\bd
+\frac{\partial}{\partial h} \left\{
D_t(\s,h) \left[h\minus\theta\minus J_0\bra\tanh(\beta H)\ket_{D_t}\right]
- \frac{J}{N\sqrt{N}} \sum_{i\neq j}\bra\bra z_{ij}
\tanh(\beta h_i(\bsigma)) \delta_{\s,\sigma_j}
\delta[h\minus h_j(\bsigma)] \ket\ket_{D_t}
\right\}
\ed
\be
 + J^2\frac{\partial^2 }{\partial h^2} \left\{
\frac{1}{N^2}\sum_{i\neq j}\bra\bra z^2_{ij}\left[1\minus\sigma_i\tanh(\beta
h_i(\bsigma)) \right]
\delta_{\s,\sigma_j} \delta[h \minus h_j (\bsigma)]
\ket\ket_{D_t} \right\}
+ {\cal O} (N^{-\frac{1}{2}})
\label{eq:closeddiffusion}
\ee
with
\be
\bra\bra f[\bsigma;\{z_{kl}\}]\ket\ket_{D}=\lim_{n\rightarrow 0}
\sum_{\bsigma^1}\cdots\sum_{\bsigma^n}
\bra f[\bsigma^1;\{z_{kl}\}]\prod_{\alpha=1}^n
\prod_{\s\mu}
\delta\left[D(\s,h_\mu)\minus
\frac{1}{N}\sum_j \delta_{\s,\sigma^\alpha_j}
\delta\left[h_\mu\minus h_j (\bsigma^\alpha) \right]\right]
\ket_{\{z_{kl}\}}
\label{eq:closedaverage}
\ee
Finally, it will become clear shortly that the diffusion term in
(\ref{eq:closedaverage}) is relatively simple, essentially
obtained by replacing $z_{ij}^2\rightarrow 1$. As a result all
complications are concentrated in a single term ${\cal A}$,
and we find for
$N\rightarrow\infty$
the relatively simple final expression
\bd
\frac{\partial}{\partial t} D_t(\s,h) =
\frac{1}{2}\left[1\plus\s\tanh(\beta h)\right]D_t(\minus\s,h)
-\frac{1}{2}\left[1\minus\s\tanh(\beta h)\right]D_t(\s,h)
\ed
\be
+\frac{\partial}{\partial h} \left\{D_t(\s,h)
\left[h\minus\theta\minus J_0\bra\tanh(\beta
H)\ket_{D_t}\right]
+ {\cal A}[\s,h;D_t]
+J^2\left[1\minus\bra\sigma\tanh(\beta H)\ket_{D_t}\right]
\frac{\partial}{\partial h}D_t(\s,h)\right\}
\label{eq:finaldiffusion}
\ee
with
\be
{\cal A}[\s,h;D_t]=-\lim_{N\rightarrow\infty}
\frac{J}{N\sqrt{N}} \sum_{i\neq j}\bra\bra z_{ij}
\tanh(\beta h_i(\bsigma)) \delta_{\s,\sigma_j}
\delta[h\minus h_j(\bsigma)] \ket\ket_{D_t}
\label{eq:flowterm}
\ee

\section{Replica Calculation of the Flow}

We now turn to the evaluation of
$\bra\bra f[\bsigma;\{z_{kl}]\ket\ket_D$.

\subsection{Disorder- and Spin-Averages}

In the familiar fashion for replica calculations we carry out the
disorder averages before the spin
averages. We remove the disorder dependence through the local
fields from within the constraining delta functions, by inserting
\bd
	1 = \prod_{\alpha} \prod_k \int \! d H^\alpha_k \ \delta
[H^\alpha_k \minus h_k ({\bsigma}^\alpha)] =
\prod_{\alpha} \prod_k \int \!
\frac{d \hat{h}^\alpha_k  d H^\alpha_k}{2 \pi}
e^{i\hat{h}^\alpha_k ( H^\alpha_k - h_k ({\bsigma}^\alpha))}
\ed
so that we can write (\ref{eq:closedaverage}) as
\bd
\bra\bra f[\bsigma;\{z_{kl}\}]\ket\ket_{D}=
\lim_{n\rightarrow 0}\int\!
\left[d\bH^1 d\hbh^1\right] \cdots
\left[d\bH^n
d\hbh^n\right]\sum_{\bsigma^1}\cdots\sum_{\bsigma^n}
e^{i\sum_i\sum_\alpha\hat{h}_i^\alpha[H_i^\alpha-\theta]
-\frac{i J_0}{N}\sum_{i\neq j}\sum_\alpha
\hat{h}^\alpha_i\sigma_j^\alpha}
\ed
\be
\prod_{\s\mu\alpha}
\delta\left[D(\s,h_\mu)\minus
\frac{1}{N}\sum_j \delta_{\s,\sigma^\alpha_j}
\delta[h_\mu\minus H^\alpha_j ]\right]~
\bra f[\bsigma^1;\{z_{kl}\}]~
e^{-\frac{iJ}{\sqrt{N}}\sum_{i\neq j}z_{ij}\sum_{\alpha}
\hat{h}^\alpha_i\sigma_j^\alpha}
\ket_{\{z_{kl}\}}
\label{eq:replicatedclosedaverage}
\ee
After symmetrisation due to $z_{ij}=z_{ji}$ and after using permutation
symmetry with respect to site labels, we find that the disorder
averages in (\ref{eq:closeddiffusion}) involve the following two
integrals,
encountered in the flow term (\ref{eq:flowterm}), and in the diffusion
term, respectively:
\bd
\int \!
\prod_{i<j}D z_{i j}e^{-\frac{iJ}{\sqrt{N}} z_{ij}\sum_{\alpha}(
\hat{h}^\alpha_i\sigma_j^\alpha+\hat{h}^\alpha_j\sigma_i^\alpha)} \left\{\!
\begin{array}{ccc} \sqrt{N} z_{12} \\ z_{12}^2
\end{array}\! \right\}
=~~~~~~~~~~~~~~~~~~~~~~~~~~~~~~~~~~~~~~~~~~~~~~~~~~~~~~~~~~~~~~~~~~~~~~
{}~~~~~~~~~~~~~~~
\ed
\be
\left\{ \! \begin{array}{ccc} \minus iJ \sum_\alpha \left[
\hat{h}^\alpha_2 \sigma^\alpha_1 \plus \hat{h}^\alpha_1 \sigma^\alpha_2
\right] \\  1 \end{array} \!\right\}\prod_{i<j}e^{
-\frac{J^2}{2N}\left[\sum_{\alpha}(\hat{h}^\alpha_i
\sigma^\alpha_j + \hat{h}^\alpha_j \sigma^\alpha_i) \right]^2}
\label{eq:disorderaverages}
\ee
with the Gaussian measure
$Dz=(2\pi)^{-\frac{1}{2}}e^{-\frac{1}{2}z^2}$, and where we only
retained the leading $\order(1)$ contributions. Applying
(\ref{eq:replicatedclosedaverage}) to the trivial function
$f[\bsigma;\{z_{kl}\}]=1$ gives a normalisation relation, which we can
use to avoid having to perform the remaining integrals. As a result we
immediately find
the diffusion term in (\ref{eq:closeddiffusion}) to be simply
\be
J^2\frac{\partial^2 }{\partial h^2} \left\{
D_t(\s,h)
\left[1\minus\bra\sigma\tanh(\beta H)\ket_{D_t}\right]
\right\}
\label{eq:diffusionterm}
\ee
which proves (\ref{eq:finaldiffusion}),
whereas the remaining disorder-induced flow term (\ref{eq:flowterm})
remains non-trivial. A similar calculation shows, upon substitution of
(\ref{eq:delta}) into (\ref{eq:macromaster}), that the second term in
the macroscopic stochastic equation (\ref{eq:macromaster}) is of order
$\ell^2 N^{-1}$. Given our assumption that the limit
$\ell\rightarrow\infty$
can be
taken after the limit $N\rightarrow\infty$, this proves
that the evolution of the
distribution $D_t(\s,h)$ is indeed deterministic on finite time-scales.

We now introduce the following set of order parameters (with their
conjugates) in order to a achieve a factorisation over sites of
(\ref{eq:flowterm}), by inserting appropriate integral representations
of unity (from which all factors $2\pi$ will vanish in the limit
$n\rightarrow0$):
\bd
m_\alpha(\{\bsigma\}) = \frac{1}{N} \sum_i \sigma^\alpha_i~~~~~~~~
W_{\alpha \beta} (\{\hbh,\bsigma\}) = \frac{1}{N}
\sum_i \hat{h}^\alpha_i \hat{h}^\beta_i \sigma^\alpha_i\sigma^\beta_i
\ed
\bd
q_{\alpha\beta}(\{\bsigma\})=\frac{1}{N}\sum_i\sigma^\alpha_i\sigma^\beta_i
{}~~~~~~~~
R_{\alpha \beta} (\{\hbh,\bsigma\}) = \frac{1}{N} \sum_i
\hat{h}^\alpha_i \sigma^\beta_i~~~~~~~~
Q_{\alpha \beta}(\{\hbh\})=\frac{1}{N}\sum_i
\hat{h}^\alpha_i \hat{h}^\beta_i
\ed
The $\delta$-distribution involving $D_t(\s,h)$ is also written in
integral form.
Combination of the trio
(\ref{eq:flowterm},\ref{eq:replicatedclosedaverage},\ref{eq:disorderaverages})
then leads to a fully site-factorised expression:
\bd
{\cal
A}[\s,h;D]=~~~~~~~~~~~~~~~~~~~~~~~~~~~~~~~~~~~~~~~~~~~~~~~~~~~~~~~~~~~~~
{}~~~~~~~~~~~~~~~~~~~~~~~~~~~~~~~~~~~~~~~~~~~~~~~
\ed
\bd
i J^2 \lim_{N\rightarrow\infty}
\lim_{n\rightarrow 0}
\int\!d\bm d\hbm d\bq d\hbq d\bQ d\hbQ d\bR d\hbR d\bW d\hbW d\hbD~
e^{J^2\sum_{\alpha\beta}W_{\alpha\beta}-\frac{N}{2}J^2
\sum_{\alpha\beta}
\left[Q_{\alpha\beta}q_{\alpha\beta}
+R_{\alpha\beta}R_{\beta\alpha}\right]}
\ed
\bd
e^{iN\sum_\alpha\left[\sum_{\s^{\prime}\mu}D(\s^{\prime},h_\mu)
\hat{D}_\alpha(\s^\prime,h_\mu)
+ m_\alpha \hat{m}_\alpha\right]
+iN\sum_{\alpha\beta} \left[
q_{\alpha\beta}\hat{q}_{\alpha\beta}+Q_{\alpha\beta}\hat{Q}_{\alpha\beta}
+R_{\alpha\beta}\hat{R}_{\alpha\beta}
+W_{\alpha\beta}\hat{W}_{\alpha\beta}\right]}
\ed
\bd
\int\!\left[d\bH^1 d\hbh^1\right] \cdots
\left[d\bH^n
d\hbh^n\right]\sum_{\bsigma^1}\cdots\sum_{\bsigma^1}
\tanh(\beta H^1_1)
\delta[h\minus H^1_2]
\delta_{\s,\sigma^1_2}
\sum_\alpha
\left[\hat{h}^\alpha_2 \sigma^\alpha_1 \plus \hat{h}^\alpha_1 \sigma^\alpha_2
\right]
\ed
\bd
e^{-i\sum_i\sum_{\alpha\beta} \left[
\hat{q}_{\alpha\beta}\sigma^\alpha_i\sigma^\beta_i
+\hat{Q}_{\alpha\beta}\hat{h}^\alpha_i \hat{h}^\beta_i
+\hat{R}_{\alpha\beta}\hat{h}^\alpha_i \sigma^\beta_i
+\hat{W}_{\alpha\beta}\hat{h}^\alpha_i \hat{h}^\beta_i
\sigma^\alpha_i\sigma^\beta_i\right]}
\ed
\bd
e^{-i\sum_i\sum_\alpha\left[
\sum_{\mu}\hat{D}_{\alpha}(\sigma_i^\alpha,h_\mu)
\delta[h_\mu\minus H^\alpha_i]
+\hat{m}_\alpha\sigma^\alpha_i
-\hat{h}_i^\alpha[H_i^\alpha-\theta- J_0 m_\alpha
+\frac{J_0}{N}\sigma_i^\alpha]\right]}
\ed
\bd
=i J^2 \lim_{N\rightarrow\infty}
\lim_{n\rightarrow 0}
\int\!d\bm d\hbm d\bq d\hbq d\bQ d\hbQ d\bR d\hbR d\bW d\hbW d\hbD~
e^{N\Psi+\order(1)}
\ed
\bd
\sum_\alpha
\left\{
\bra \tanh(\beta H_1)\sigma_\alpha \ket_M
\bra \delta[h\minus H_1]\delta_{\s,\sigma_1}\hat{h}_\alpha\ket_M
+\bra\tanh(\beta H_1) \hat{h}_\alpha\ket_M
\bra \delta[h\minus H_1]\delta_{\s,\sigma_1}\sigma_\alpha\ket_M
\right\}
\ed
with
\bd
\Psi=
i\sum_{\alpha\sigma\mu}D(\sigma,h_\mu)\hat{D}_\alpha(\sigma,h_\mu)
+ i\sum_\alpha m_\alpha \hat{m}_\alpha
+i\sum_{\alpha\beta} \left[
q_{\alpha\beta}\hat{q}_{\alpha\beta}\plus
Q_{\alpha\beta}\hat{Q}_{\alpha\beta}
\plus R_{\alpha\beta}\hat{R}_{\alpha\beta}\plus
W_{\alpha\beta}\hat{W}_{\alpha\beta}\right]
\ed
\bd
-\frac{1}{2}J^2 \sum_{\alpha\beta}\left[Q_{\alpha\beta}q_{\alpha\beta}
+R_{\alpha\beta}R_{\beta\alpha}\right]
+\log \int\!d\bH d\hbh\bra M[\bH,\hbh,\bsigma]\ket_{\bsigma}
\ed
(in which $\bra f(\bsigma)\ket_{\bsigma}=
2^{-n}\sum_{\sigma_1}\cdots\sum_{\sigma_n}f(\bsigma)$)
and with the effective single-site measure $M$ (all vectors
now carry replica-indices only):
\be
\bra f[\bH,\hbh,\bsigma] \ket_M=
\frac{\int\!d\bH d\hbh \sum_{\bsigma}
M[\bH,\hbh,\bsigma]f[\bH,\hbh,\bsigma]}
{\int\!d\bH d\hbh\sum_{\bsigma} M[\bH,\hbh,\bsigma]}
\label{eq:effectivemeasure}
\ee
\bd
M[\bH,\hbh,\bsigma]=\exp\left\{
\minus i\hbm\cdot\bsigma \minus i\bsigma\cdot\hbq\bsigma
\minus i \hbh\cdot\hbQ\hbh \minus i \hbh\cdot\hbR\bsigma
\minus i \sum_{\alpha\beta}\hat{W}_{\alpha\beta}\hat{h}_\alpha\hat{h}_\beta
\sigma_\alpha\sigma_\beta
{}~~~~~~~~
\right.
\ed
\bd
\left.
{}~~~~~~~~~~~~~~~~~~~~~~~~~~~~~~~~~~~~
- i\sum_{\alpha\mu}\hat{D}_{\alpha}(\sigma_\alpha,h_\mu)
\delta[h_\mu\minus H_\alpha]
\plus i\sum_\alpha\hat{h}_\alpha[H_\alpha\minus \theta\minus
J_0 m_\alpha]
\right\}
\ed
By changing the order of the limits $N\rightarrow\infty$ and
$n\rightarrow0$, the remaining integral can be evaluated by steepest
descent. It is dominated by the extremum of $\Psi$ which for $n>1$
defines a global maximum
(the $\order(1)$ term in the exponent will drop out due to
normalisation,
as can be checked
explicitly by using the above calculation to rewrite $\bra\bra 1\ket\ket_D$).

\subsection{Simplification of the Saddle-Point Problem}

We can make several immediate simplifications. Firstly, variation of $\Psi$
with respect to $W_{\alpha\beta}$, $Q_{\alpha\beta}$,
$q_{\alpha\beta}$ and $R_{\alpha\beta}$ gives saddle-point equations
with which to remove
all conjugate order parameter matrices from our
problem:
\bd
\hbW=0~~~~~~~~i\hbQ=\frac{1}{2}J^2\bq~~~~~~~~i\hbq=\frac{1}{2}J^2\bQ~~~~~~~~
i\hbR=J^2 \bR^{\dag}
\ed
Secondly, the scaling freedom in the definition of
the conjugate parameters $\hat{D}(\s^\prime,h_\mu)$ can be used to take the
limit $\ell\rightarrow\infty$:
\bd
\sum_{\mu}\hat{D}_\alpha(\sigma,h_\mu)f(h_\mu)\rightarrow
\sum_{\mu}\Delta h.\hat{D}_\alpha(\sigma,h_\mu)f(h_\mu)\rightarrow
\int\!dH~\hat{D}_\alpha(\sigma,H)f(H)~~~~~~(\ell\rightarrow\infty)
\ed
The result of these simplications and of taking the
$N\rightarrow\infty$ limit is the following:
\bd
{\cal A}[\s,h;D]=i J^2 \lim_{n\rightarrow
0}~~~~~~~~~~~~~~~~~~~~~~~~~~~~~~~~~~~~~~~~~~~~~~~~~~~~~~~~~~~~~~~~~~
{}~~~~~~~~~~~~~~~~~~~~~~~~~~~~~~~~~~~~~~~
\ed
\bd
\sum_\alpha
\left\{
\bra \tanh(\beta H_1)\sigma_\alpha \ket_M
\bra \delta[h\minus H_1]\delta_{\s,\sigma_1}\hat{h}_\alpha\ket_M
+\bra\tanh(\beta H_1) \hat{h}_\alpha\ket_M
\bra \delta[h\minus H_1]\delta_{\s,\sigma_1}\sigma_\alpha\ket_M
\right\}
\ed
with the effective measure (\ref{eq:effectivemeasure}), in which $M$
and the exponent $\Psi$ to be extremised are now given by
\bd
M[\bH,\hbh,\bsigma]=e^{
- i\hbm\cdot\bsigma - \frac{1}{2}J^2\bsigma\cdot\bQ\bsigma
- \frac{1}{2}J^2\hbh\cdot\bq\hbh
-i\sum_{\alpha}\hat{D}_{\alpha}(\sigma_\alpha,H_\alpha)
+ i\hbh\cdot[\bH\minus \btheta\minus J_0 \bm\plus iJ^2\bR^{\dag}\bsigma]}
\ed
\bd
\Psi=i\sum_{\alpha\sigma}\int\!dH~D(\sigma,H)\hat{D}_\alpha(\sigma,H)
+ i\sum_\alpha m_\alpha \hat{m}_\alpha
+\frac{1}{2}J^2\sum_{\alpha\beta} \left[
q_{\alpha\beta}Q_{\alpha\beta}\plus R_{\alpha\beta}R_{\beta\alpha}
\right]
\ed
\bd
+\log \int\!d\bH d\hbh\bra M[\bH,\hbh,\bsigma]\ket_{\bsigma}
\ed
with the notation $\btheta=(\theta,\ldots,\theta)$.
Next we perform the integrations over the conjugate fields $\hbh$,
which leads to an effective measure $M$ involving spins and fields
only:
\be
\bra f[\bH,\bsigma] \ket_M=
\frac{\int\!d\bH\sum_{\bsigma}
M[\bH,\bsigma]f[\bH,\bsigma]}
{\int\!d\bH\sum_{\bsigma} M[\bH,\bsigma]}
\label{eq:rsbmeasure}
\ee
\be
M[\bH,\bsigma]=e^{
- i\hbm\cdot\bsigma - \frac{1}{2}J^2\bsigma\cdot\bQ\bsigma
-i\sum_{\alpha}\hat{D}_{\alpha}(\sigma_\alpha,H_\alpha)
-\frac{1}{2J^2}
[\bH\minus \btheta\minus J_0 \bm\plus iJ^2\bR^{\dag}\bsigma]\cdot\bq^{-1}
[\bH\minus \btheta\minus J_0 \bm\plus iJ^2\bR^{\dag}\bsigma]
}
\label{eq:rsbM}
\ee
\bd
\Psi=
i\sum_{\alpha\sigma}\int\!dH~D(\sigma,H)\hat{D}_\alpha(\sigma,H)
+ i\sum_\alpha m_\alpha \hat{m}_\alpha
+\frac{1}{2}J^2\sum_{\alpha\beta} \left[
q_{\alpha\beta}Q_{\alpha\beta}\plus R_{\alpha\beta}R_{\beta\alpha}
\right]
\ed
\be
-\frac{1}{2}\log{\rm det}~\bq
+\log \int\!d\bH \bra M[\bH,\bsigma]\ket_{\bsigma}
\label{eq:rsbpsi}
\ee
In $\Psi$ (\ref{eq:rsbpsi}) we have neglected irrelevant constants.
At this stage it will be convenient to calculate the remaining
saddle-point equations, by variation of (\ref{eq:rsbpsi}). The first
of these equations, obtained by variation
with respect to $\hat{D}(\sigma,H)$,
enables us to write all averages with a single replica-index, involving
fields and spins, selfconsistently in terms of the original
distribution $D(\sigma,H)$:
\be
D(\sigma,H)=\bra \delta_{\sigma,\sigma_{\alpha}}\delta[H\minus H_\alpha]
\ket_M
\label{eq:rsbsaddleD}
\ee
\be
m_\alpha=m=\bra \sigma\ket_D
\label{eq:rsbsaddlem}
\ee
\be
q_{\alpha\beta}=\bra \sigma_\alpha \sigma_\beta \ket_M
\label{eq:rsbsaddleq}
\ee
\be
\hat{m}_\alpha = i\frac{J_0}{J^2}
\sum_{\beta}(q^{-1})_{\alpha\beta}\left\{
\bra H\ket_D
- \theta- J_0 m +
iJ^2 m\sum_\gamma R_{\gamma\beta}\right\}
\label{eq:rsbhatm}
\ee
\be
R_{\alpha\beta}=\frac{i}{J^2}\sum_{\gamma}
(q^{-1})_{\alpha\gamma}
\bra [H_\gamma\minus \theta\minus J_0 m \plus iJ^2(\bR^{\dag}\bsigma)_\gamma]
\sigma_\beta
\ket_M
\label{eq:rsbR}
\ee
\be
J^2 Q_{\alpha\beta} =
\frac{\partial}{\partial q_{\alpha\beta}}\log{\rm det}~\bq
- 2 \bra \frac{\partial \log M[\bH,\bsigma]}{\partial q_{\alpha\beta}}\ket_M
\label{eq:rsbQ}
\ee
We can now write the flow term $\cal A$ (\ref{eq:flowterm}) of
our diffusion equation as
\bd
{\cal A}[\s,h;D]=-\lim_{n\rightarrow 0}\sum_{\alpha\beta}(q^{-1})_{\alpha\beta}
{}~~~~~~~~~~~~~~~~~~~~~~~~~~~~~~~~~~~~~~~~~~~~~~~~~~~~~~~~~~~~~~~~~~~~~~~~
{}~~~~~~~~~~~~~~~~~~
\ed
\bd
\left\{~
\bra \tanh(\beta H_1)\sigma_\alpha \ket_M ~
\bra \delta[h\minus H_1]\delta_{\s,\sigma_1}
[H_\beta\minus \theta\minus J_0 m\plus iJ^2(\bR^{\dag}\bsigma)_\beta]
\ket_M~~~~~~
\right.
\ed
\be
\left.
{}~~~~~~+~
\bra \delta[h\minus H_1]\delta_{\s,\sigma_1}\sigma_\alpha\ket_M~
\bra\tanh(\beta H_1)
[H_\beta\minus \theta\minus J_0 m\plus iJ^2(\bR^{\dag}\bsigma)_\beta]
\ket_M~
\right\}
\label{eq:rsbintensivepart}
\ee

\subsection{Equilibrium}

In equilibrium, we know that the microscopic probability distribution
is of the Boltzmann form, $p_\infty (\bsigma) \sim e^{- \beta
H(\bsigma)}$. Therefore, the present constraint restricting
micro-states under consideration to those with the same joint
spin-field distribution,
must in equilibrium reduce
to a constraint selecting states with the same energy.
We will now make the ansatz
\be
\hat{D}_\alpha(\sigma,H)=
\frac{1}{2}i\beta\sigma\left[H+\theta\right]
\label{eq:equilansatz}
\ee
and show that it indeed corresponds to a
stationary
state for our diffusion
equation (\ref{eq:finaldiffusion}), in which one recovers the familiar
equations from equilibrium statistical mechanics; i.e. the full (RSB)
order parameter equations \cite{SK,parisi} as well as the equilibrium
local field distribution
\cite{thomsenetal}.

We first turn to the saddle-point equations.
Given the simple expression (\ref{eq:equilansatz}) we can perform the
field integrations, with the result
\bd
\Psi=-\frac{1}{2}\beta n\sum_{\sigma}
\int\!dH~ D(\sigma,H)\sigma[H\plus \theta]
+ i\sum_\alpha m_\alpha \hat{m}_\alpha
+\frac{1}{2}J^2\sum_{\alpha\beta} \left[
q_{\alpha\beta}Q_{\alpha\beta}\plus R_{\alpha\beta}R_{\beta\alpha}
\right]
\ed
\be
+\log \bra e^{[\beta\btheta+\frac{1}{2}\beta J_0\bm-i\hat{\bm}]\cdot\bsigma
+\frac{1}{2}J^2\bsigma\cdot[\frac{1}{4}\beta^2\bq-\bQ-i\beta\bR^{\dag}]
\bsigma} \ket_{\bsigma}
\ee
(again we forget about irrelevant constants). The remaining saddle-point
equations become
\be
\hat{m}_\alpha=\frac{1}{2}i\beta J_0 m_\alpha~~~~~~~~
Q_{\alpha\beta}=-\frac{1}{4}\beta^2 q_{\alpha\beta}~~~~~~~~
R_{\alpha\beta}=\frac{1}{2}i\beta q_{\beta\alpha}
\label{eq:equilsaddle1}
\ee
\be
m_\alpha =
\frac{ \bra \sigma_\alpha e^{\beta[J_0\bm+\btheta]\cdot\bsigma
+\frac{1}{2}(\beta J)^2\bsigma\cdot\bq\bsigma}
\ket_{\bsigma}}
{\bra e^{\beta[J_0\bm+\btheta]\cdot\bsigma
+\frac{1}{2}(\beta J)^2\bsigma\cdot\bq\bsigma}
\ket_{\bsigma}}~~~~~~~~
q_{\alpha\beta}=
\frac{ \bra \sigma_\alpha\sigma_\beta e^{\beta[J_0\bm+\btheta]\cdot\bsigma
+\frac{1}{2}(\beta J)^2\bsigma\cdot\bq\bsigma}
\ket_{\bsigma}}
{\bra e^{\beta[J_0\bm+\btheta]\cdot\bsigma
+\frac{1}{2}(\beta J)^2\bsigma\cdot\bq\bsigma}
\ket_{\bsigma}}
\label{eq:equilsaddle2}
\ee
which are the familiar equations \cite{parisi} as obtained by an
equilibrium (thermodynamic) analysis.
With the relations (\ref{eq:equilansatz},\ref{eq:equilsaddle1}) we can simplify
the effective measure $M$ considerably:
\be
\bra f[\bH,\bsigma] \ket_M=
\frac{\int\!d\bz~e^{-\frac{1}{2}\bz\cdot\bq^{-1}\bz}
\bra f[J_0\bm\plus \btheta\plus J\bz,\bsigma]
e^{\beta\bsigma[J_0\bm+\btheta+J\bz]} \ket_{\bsigma}}
{\int\!d\bz~e^{-\frac{1}{2}\bz\cdot\bq^{-1}\bz}\bra
e^{\beta\bsigma[J_0\bm+\btheta+J\bz]} \ket_{\bsigma}}
\label{eq:simplemeasure}
\ee
(with $\bm=(m,\ldots,m)$)
This simplified measure obeys useful relations like
\be
\bra \sigma_\alpha f[\bH;\{\sigma_{\gamma\neq\alpha}\}] \ket_M =
\bra \tanh(\beta H_\alpha) f[\bH;\{\sigma_{\gamma\neq\alpha}\}] \ket_M
\label{eq:nicerelation1}
\ee
\be
\bra(H_\alpha\minus
J_0m\minus\theta)f[\{H_{\gamma\neq\alpha}\};\bsigma]\ket_M
=
\beta J^2 \sum_{\beta}q_{\alpha\beta}\bra \sigma_\beta
f[\{H_{\gamma\neq\alpha}\};\bsigma]\ket_M
\label{eq:nicerelation2}
\ee
In particular we now find $m=\bra \tanh(\beta H)\ket_D$.
If we combine the expression (\ref{eq:simplemeasure}) with
(\ref{eq:rsbsaddleD}),
sum over the remaining spin
variable $\sigma$ and perform the integration over $\bz$, we are led
directly to the
equilibrium expression for the local field distribution as obtained
in \cite{thomsenetal}:
\bd
D(h)=\lim_{n\rightarrow0}\int\frac{dk}{2\pi}e^{-\frac{1}{2}J^2
k^2-ik(h-J_0m-\theta)}
\frac{\bra e^{\beta[J_0\bm\plus\btheta]\cdot\bsigma+\frac{1}{2}(\beta
J)^2\bsigma\cdot\bq\bsigma+ik\beta
J^2\sum_{\alpha}q_{1\alpha}\sigma_\alpha}
\ket_{\bsigma}}
{\bra e^{\beta[J_0\bm\plus\btheta]\cdot\bsigma+\frac{1}{2}(\beta
J)^2\bsigma\cdot\bq\bsigma}\ket_{\bsigma}}
\ed
\vsp

Next we show that the choice (\ref{eq:equilansatz}) corresponds
to a fixed point of the diffusion equation (\ref{eq:finaldiffusion}),
i.e. that   $\frac{d }{d t} D_t(\varsigma,h) = 0$ for all
$(\varsigma,h)$.
In the right-hand side of
(\ref{eq:finaldiffusion}) the first two terms trivially cancel, which
follows from applying to (\ref{eq:rsbsaddleD}) the identities
$\delta_{\s,\sigma}=\frac{1}{2}[1\plus\s\sigma]$  and
(\ref{eq:nicerelation1}):
\bd
[1\plus\s\tanh(\beta h)]D(\minus\s,h)-
[1\minus\s\tanh(\beta h)]D(\s,h)=
\s\bra \delta[h\minus H_\alpha]\left[\tanh(\beta
h)\minus\sigma_\alpha\right]\ket_M=0
\ed
Equivalently:
\bd
D(\s,h)=\frac{1}{2}[1\plus \s\tanh(\beta h)]D(h)
\ed
We use (\ref{eq:equilsaddle1}) and
(\ref{eq:nicerelation1},\ref{eq:nicerelation2}) to rewrite
(\ref{eq:rsbintensivepart}). In doing so we will also use equilibrium
relations like
\bd
\beta J^2\sum_{\alpha}q_{1\alpha}^2=\bra\tanh(\beta H)(H\minus
J_0m\minus\theta)\ket_D
\ed
which can be derived directly from the equilibrium saddle-point
equations (see e.g. \cite{sk1}). The result is:
\bd
{\cal A}[\s,h;D]= -(h\minus \theta\minus J_0 m)D(\s,h)
-\beta J^2[1\minus\bra\tanh^2(\beta H)\ket_D]\s D(\s.h)~~~~~~~~~~
\ed
\be
{}~~~~~~~~~~
+[1\minus\bra\tanh^2(\beta H)\ket_D]
\lim_{n\rightarrow0}\sum_\gamma (q^{-1})_{1\gamma}
\bra \delta[h\minus H_1]\delta_{\s,\sigma_1}[H_\gamma\minus
J_0 m\minus\theta]\ket_M
\label{eq:equilflowterm}
\ee
In order to combine the flow term ${\cal A}$ in (\ref{eq:finaldiffusion})
with the diffusion term, we apply (\ref{eq:simplemeasure}) to equation
(\ref{eq:rsbsaddleD}) and calculate the field derivative:
\bd
J^2\frac{\partial}{\partial h}D(\s,h)= J\lim_{n\rightarrow0}
\frac{\int\!d\bz~\delta[h\minus J_0 m\minus \theta\minus
Jz_\alpha]\frac{\partial}{\partial z_\alpha}\left\{
e^{-\frac{1}{2}\bz\cdot\bq^{-1}\bz}
\bra
\delta_{\s,\sigma_\alpha}
e^{\beta\bsigma[J_0\bm+\btheta+J\bz]} \ket_{\bsigma}\right\}}
{\int\!d\bz~e^{-\frac{1}{2}\bz\cdot\bq^{-1}\bz}\bra
e^{\beta\bsigma[J_0\bm+\btheta+J\bz]} \ket_{\bsigma}}
\ed
\be
=
\beta J^2\s D(\s,h)-\lim_{n\rightarrow0}
\sum_\gamma (q^{-1})_{1\gamma}
\bra \delta[h\minus H_1]\delta_{\s,\sigma_1}[H_\gamma\minus
J_0 m\minus\theta]\ket_M
\label{eq:equildiffusion}
\ee
Insertion of (\ref{eq:equilflowterm}) and (\ref{eq:equildiffusion})
into the right-hand side of (\ref{eq:finaldiffusion}) leads to the
desired result: it exactly
vanishes. This completes the proof that the standard thermodynamic
equilibrium state, as calculated within equilibrium statistical
mechanics, defines a fixed-point of our diffusion equation
(\ref{eq:finaldiffusion}). Note, however, that this leaves open the
possibility of existence for stationary states other than the
thermodynamic one.

\section{Replica Symmetric Flow}

\subsection{Derivation of the RS Equations}

In order to proceed further in evaluating explicitly the saddle points we now
make, as a first step, the ergodicity or replica-symmetry ansatz
(RS).
All order
parameters with a single replica index are assumed not to depend on
this index; all order parameter matrices are assumed to have entries
which depend only on whether or not they are on the diagonal. With a
modest amount of foresight we put:
\be
\begin{array}{lll}
\room m_\alpha=m & &
q_{\alpha\beta}=(1\minus q)\delta_{\alpha\beta}\plus q \\
\room \hat{m}_\alpha=i\mu & &
R_{\alpha\beta}=i(1\minus q)[R_0\delta_{\alpha\beta}\plus R] \\
\room \hat{D}_\alpha(\sigma,H)=i\log \chi(\sigma,H)& &
Q_{\alpha\beta}=Q_0\delta_{\alpha\beta}\plus qR_0^2\minus
2(1\minus q)RR_0\minus Q^2
\end{array}
\label{eq:RSansatz}
\ee
which implies $(q^{-1})_{\alpha\beta}=(1\minus
q)^{-1}[\delta_{\alpha\beta}\minus q(1\minus q)^{-1}]\plus\order(n)$.
Working out the RS version of the extensive exponent $\Psi$ (\ref{eq:rsbpsi})
gives:
\bd
\lim_{n\rightarrow0}\frac{\Psi_{\rm RS}}{n}=
-m\mu -\frac{1}{2}\log(1\minus q)-\frac{q}{2(1\minus q)}
-\frac{1}{2}J^2(1\minus q)Q^2-J^2(1\minus q)^2[R_0^2\plus 2R_0R]
\ed
\be
-\sum_{\sigma}\int\!dH~D(\sigma,H)\log\chi(\sigma,H)
+\lim_{n\rightarrow0}\frac{1}{n}\log \int\!d\bH \bra
M_{\rm RS}[\bH,\bsigma]\ket_{\bsigma}
\label{eq:rsansatzinpsi}
\ee
with
\bd
M_{\rm RS}[\bH,\bsigma]=
\prod_\alpha\left\{
\chi(\sigma_\alpha,H_\alpha)
e^{
\mu\sigma_\alpha
-\frac{1}{2J^2(1-q)}(H_\alpha\minus\theta\minus J_0 m)^2
+R_0(H_\alpha\minus\theta\minus J_0 m)\sigma_\alpha}
\right\}
\ed
\be
e^{\frac{q}{2J^2(1-q)^2}[\sum_\alpha(H_\alpha\minus\theta\minus J_0 m)]^2
+\frac{1}{2}J^2 Q^2[\sum_\alpha\sigma_\alpha]^2
+(R-\frac{qR_0}{1-q})[\sum_\alpha(H_\alpha\minus\theta\minus J_0 m)]
[\sum_\beta\sigma_\beta]}
\label{eq:Mnotfactorised}
\ee
We can obtain a factorisation of $M_{\rm RS}[\bH,\bsigma]$ with
respect
to the replica
labels
by introducing appropriate Gaussian integrations:
\bd
\room
e^{A[\sum_\alpha F_\alpha]^2+B[\sum_\alpha\sigma_\alpha]^2
+C[\sum_\alpha F_\alpha][\sum_\beta\sigma_\beta]}
=
\int\!Dx
Dy\prod_\alpha
e^{F_\alpha\sqrt{2A}(x\cos\phi+y\sin\phi)
+\sigma_\alpha\sqrt{2B}(x\cos\phi-y\sin\phi)}
\ed
with $\cos(2\phi)=C/2\sqrt{AB}$.
Application the above identity to (\ref{eq:Mnotfactorised}) leads to
an expression for
(\ref{eq:rsansatzinpsi}) in
which we can take the remaining limit $n\rightarrow0$. We use the
definition of the angle $\phi$ to eliminate the order parameter $R$
from our problem and write the averages over the two Gaussian
variables
$x$ and $y$ as
$\bras \ldots \kets$. The final result involves an effective measure $M_{\rm
RS}[H,\sigma]$ without
replica indices:
\bd
\lim_{n\rightarrow0}\frac{\Psi_{\rm RS}}{n}=
-m\mu -\frac{1}{2}\log(1\minus q)-\frac{q}{2(1\minus q)}
-J^2(1\minus q)\left[
\frac{1}{2}Q^2\plus (1\plus q)R_0^2
\plus 2 R_0 Q\sqrt{q}\cos(2\phi)\right]
\ed
\be
-\sum_{\sigma}\int\!dH~D(\sigma,H)\log\chi(\sigma,H)
+\bigbras \log
\int\!dH \bra M_{\rm RS}[H,\sigma]\ket_\sigma\bigkets
\label{eq:rspsi}
\ee
with
\be
M_{\rm RS}[H,\sigma]=
\chi(\sigma,H) e^{\mu\sigma
-\frac{(H\minus\theta\minus J_0 m)^2}{2J^2(1-q)}
+R_0(H\minus\theta\minus J_0m)\sigma
+\frac{\sqrt{q}}{J(1-q)}(H\minus\theta\minus
J_0m)(x\cos\phi+y\sin\phi)+JQ\sigma(x\cos\phi-y\sin\phi)}
\label{eq:rsmeasure}
\ee
We write averages with respect to this final measure $M_{\rm
RS}[H,\sigma]$,
which are parametrised by the Gaussian variables $x$ and $y$, as
\bd
\bra f[H,\sigma]\ket_{\star}=
\frac{\int\!dH\sum_\sigma M_{\rm RS}[H,\sigma]f[H,\sigma]}
{\int\!dH\sum_\sigma M_{\rm RS}[H,\sigma]}
\ed
To further reduce our future bookkeeping we derive two useful relations by
partial integration over the Gaussian variables:
\bd
\bigbras x \bra f[H,\sigma]\ket_\star \bigkets=
\frac{\cos\phi\sqrt{q}}{J(1\minus q)}\bigbras
\bra f[H,\sigma]H\ket_\star \minus
\bra f[H,\sigma]\ket_\star \bra H\ket_\star
\bigkets
\ed
\bd
{}~~~~~~~~~~~~~~~~~~~~~~~~
+~JQ\cos\phi\bigbras \bra f[H,\sigma]\sigma\ket_\star\minus
\bra f[H,\sigma]\ket_\star \bra \sigma\ket_\star
\bigkets
\ed
\bd
\bigbras y \bra f[H,\sigma]\ket_\star \bigkets=
\frac{\sin\phi\sqrt{q}}{J(1\minus q)}\bigbras
\bra f[H,\sigma] H\ket_\star \minus
\bra f[H,\sigma]\ket_\star \bra H\ket_\star
\bigkets
\ed
\bd
{}~~~~~~~~~~~~~~~~~~~~~~~~
-~JQ\sin\phi\bigbras \bra f[H,\sigma]\sigma\ket_\star\minus
\bra f[H,\sigma]\ket_\star \bra \sigma\ket_\star
\bigkets
\ed
Functional differentiation of (\ref{eq:rspsi})
with respect to the
function $\chi$ gives the RS saddle-point equation
\be
D(\s,h)=\bigbras\bra \delta[h\minus
H]\delta_{\s,\sigma}\ket_\star\bigkets
\label{eq:RS1}
\ee
which implies, as expected, the general relation
\bd
\bigbras \bra f[H,\sigma]\ket_{\star}\bigkets =\bra f[H,\sigma]\ket_D
\ed
Differentiation of (\ref{eq:rspsi})
with respect to the parameters $\{q,m,\mu,R_0,Q,\phi\}$
and repeated usage of the above bookkeeping identities gives the
remaining
RS saddle-point equations:
\be
m= \bra \sigma \ket_D
\label{eq:RS2}
\ee
\be
q=\bigbras \bra \sigma\ket^2_\star \bigkets
\label{eq:RS3}
\ee
\be
2J^2 R_0(1\minus q)^2= \bra \sigma (H\minus J_0m\minus\theta)\ket_D
- \bigbras \bra \sigma\ket_\star\bra H\minus J_0m\minus\theta
\ket_\star\bigkets
\label{eq:RS4}
\ee
\be
2J^2 (1\minus q)\left[R_0(1\plus q)\plus Q\sqrt{q}\cos(2\phi)\right]
=\bra \sigma(H\minus J_0m\minus\theta)\ket_D
\label{eq:RS5}
\ee
\be
\mu +J_0 R_0 m=\frac{J_0}{J^2(1\minus q)}\bra H\minus
J_0m\minus\theta\ket_D
\label{eq:RS6}
\ee
\bd
J^2 q-J^4(1\minus q)^2\left[Q^2\plus 4qR_0^2\plus
8QR_0\sqrt{q}\cos(2\phi)\right]
+\bra (H\minus J_0m\minus\theta)^2\ket_D
\ed
\be
=\frac{1\plus q}{1\minus q}\bigbras \bra (H\minus J_0m\minus\theta)^2
\ket_\star\minus \bra
H\minus J_0m\minus\theta\ket_\star^2\bigkets
\label{eq:RS7}
\ee

We now use the RS ansatz to perform the $n\rightarrow0$ limit in the
flow
term ${\cal A}$ (\ref{eq:rsbintensivepart}) of our diffusion equation
(\ref{eq:finaldiffusion}). Note that, due to $n\rightarrow0$, we may
deal with averages over the original measure $M$ which
involve two replica indices (such as those encountered in
(\ref{eq:rsbintensivepart}) in the following way:
\bd
\bra f[H_\alpha,\sigma_\alpha]g[H_\beta,\sigma_\beta]\ket_M\rightarrow
\delta_{\alpha\beta}\bra
f[H,\sigma]g[H,\sigma]\ket_D
+(1\minus\delta_{\alpha\beta})\bigbras
\bra f[H,\sigma]\ket_\star\bra g[H,\sigma]\ket_\star\bigkets
\ed
With this identity we can work out (\ref{eq:rsbintensivepart}). We use
the short-hand $Q\sqrt{q}\cos(2\phi)=(1\minus q)R_1$,
and find after some bookkeeping and some re-arranging of terms:
\bd
(1\minus q)^2 {\cal A}_{\rm RS}[\s,h;D]=
(2q\minus 1)D(\s,h)\left[\room
(h\minus J_0m\minus\theta)\bra \tanh(\beta H)\sigma\ket_D
\plus\s\bra\tanh(\beta H)(H\minus J_0m\minus\theta)\ket_D\right]
\ed
\bd
-q D(\s,h)\left[
(h\minus J_0m\minus\theta)
\bigbras\bra\tanh(\beta H)\ket_\star\bra\sigma\ket_\star\bigkets \plus
\s \bigbras\bra\tanh(\beta H)\ket_\star\bra H\minus J_0m\minus\theta
\ket_\star\bigkets
\right]
\ed
\bd
+2\s J^2(1\minus q)^2 D(\s,h)\left[(R_1\plus R_0)\bra\tanh(\beta H)
\sigma\ket_D\minus
R_1\bigbras \bra\tanh(\beta H)\ket_\star\bra\sigma\ket_\star\bigkets
\right]
\ed
\bd
+\bigbras\bra\delta[h\minus H]\delta_{\s,\sigma}\ket_\star\bra H\minus
J_0m\minus\theta\ket_\star\bigkets
\left[
\bigbras\bra\tanh(\beta H)\ket_\star\bra\sigma\ket_\star\bigkets\minus
q\bra\tanh(\beta H)\sigma\ket_D\right]
\ed
\bd
+\bigbras\bra\delta[h\minus H]\delta_{\s,\sigma}\ket_\star\bra\sigma
\ket_\star\bigkets
\left[
\bigbras\bra\tanh(\beta H)\ket_\star\bra H\minus
J_0m\minus\theta\ket_\star\bigkets\minus
q\bra\tanh(\beta H)(H\minus J_0m\minus\theta)\ket_D\right]
\ed
\be
+2 J^2(1\minus q)^2\bigbras\bra\delta[h\minus
H]\delta_{\s,\sigma}\ket_\star\bra\sigma\ket_\star\bigkets
 \left[(R_1\minus R_0)
\bigbras\bra\tanh(\beta H)\ket_\star\bra\sigma\ket_\star\bigkets \minus
R_1\bra\tanh(\beta H)\sigma\ket_D
\right]
\label{eq:ooff}
\ee
In RS approximation the evolution of
the joint spin-field distribution is described by equation
(\ref{eq:finaldiffusion}), in which the disorder-induced term ${\cal
A}$
is given
by (\ref{eq:ooff}). Evaluation of ${\cal A}$, in turn, requires
solving the set of saddle-point equations (\ref{eq:RS1}-\ref{eq:RS7}),
at each instance of time.

\subsection{The AT Instability}

The de Almeida-Thouless (AT) instability \cite{AT} marks the
instability of the RS solution of the saddle-point equations to
the so-called replicon mode. This leads to a second order transition
away from the RS state
to states with broken replica symmetry (RSB). Unlike the standard
equilibrium calculations, we here
have to worry about replicon fluctuations with respect to three replica
matrices:
\bea
q_{\alpha \beta} \rightarrow q^{\rm RS}_{\alpha\beta} + \delta
q_{\alpha \beta}~~
& ~~~~~~~~\delta q_{\alpha\alpha}=0
& ~~~~~~\sum_{\alpha}\delta q_{\alpha \beta} =
\sum_{\beta}\delta q_{\alpha\beta}=0
\nonumber \\
Q_{\alpha\beta}\rightarrow Q^{\rm RS}_{\alpha\beta}+\delta
Q_{\alpha\beta}
& ~~~~~~~~\delta Q_{\alpha\alpha}=0
& ~~~~~~\sum_{\alpha}\delta Q_{\alpha \beta} =
\sum_{\beta}\delta Q_{\alpha\beta}=0
 \\
{}~R_{\alpha\beta}\rightarrow R^{\rm RS}_{\alpha\beta}+i\delta
R_{\alpha\beta}
& ~~~~~~~~\delta R_{\alpha\alpha}=0
& ~~~~~~\sum_{\alpha}\delta R_{\alpha \beta} =
\sum_{\beta}\delta R_{\alpha\beta}=0
\nonumber
\label{eq:replicon}
\eea
with $\delta q_{\alpha\beta}=\delta q_{\beta\alpha}$,  $\delta
Q_{\alpha\beta}=\delta Q_{\beta\alpha}$ and $\delta
R_{\alpha\beta}=\delta R_{\beta\alpha}$.
As usual the replicon fluctuations  satisfy  convenient matrix
commutation
relations,
like $[\bq_{\rm RS},\delta\bq]=[\bQ_{\rm RS},\delta\bQ]=[\bR_{\rm
RS},\delta\bR]=0$. The AT
instability corresponds to a zero eigenvalue in the spectrum of the
Hessian (i.e. the matrix
of second derivatives) of $\Psi$ at the RS saddle-point.
However, since the $R_{\alpha\beta}$ are conjugate order
parameters, acquiring an imaginary value, the naive picture of this
zero
eigenvalue signalling
the bifurcation of a local maximum, need not be true. We can avoid
all such subtleties by following the
alternative procedure: to consider RSB fluctuations only after
elimination of the conjugate order parameters $R_{\alpha\beta}$ with
equation (\ref{eq:rsbR}). This is equivalent to first working out
the variation in $\Psi$ (\ref{eq:rsbpsi}) for the case where all
fluctuations (\ref{eq:replicon}) are independent, followed by a
projection onto the subspace defined by (\ref{eq:rsbR}).

Expansion of (\ref{eq:rsbpsi}) around the RS
saddle point, the first non-trivial order of which must by definition
be quadratic in the replicon fluctuations, gives
\bd
\Psi-\Psi_{\rm RS}=
\frac{1}{2}\bra G^2\ket_{M}
+\frac{1}{2}J^2\sum_{\alpha\beta}
\delta q_{\alpha\beta}\delta Q_{\alpha\beta}
-J^2\sum_{\alpha\beta}\delta R^2_{\alpha\beta}
-J^2 R_0\sum_{\alpha\beta}\delta R_{\alpha\beta}\delta q_{\alpha\beta}
\ed
\be
+\frac{1}{2}\sum_{\alpha\beta}\delta q^2_{\alpha\beta}\left[
\frac{1}{2(1\minus q)^2}\plus 3J^2 R_0^2\minus
\frac{1}{J^2(1\minus q)^3}\bigbras \bra H^2\ket_\star\minus \bra
H\ket^2_\star\bigkets\right]+\order(\delta^3)
\label{eq:repliconshift}
\ee
with
\bd
G=
-\frac{1}{2}J^2\bsigma\cdot\left[\delta\bQ\plus 2R_0\delta\bR
\minus R_0^2\delta\bq\right]\bsigma
+\frac{(\bH\minus J_0\bm\minus\btheta)\cdot\delta\bq(\bH\minus
J_0\bm\minus\btheta)}{2J^2(1\minus q)^2}
\ed
\bd
+\frac{\bsigma\cdot\left[\delta\bR\minus R_0\delta\bq\right](\bH\minus
J_0\bm\minus\btheta)}{1\minus q}
\ed
In order to evaluate the term in (\ref{eq:repliconshift}) that
involves $G$,
we note that in RS
saddle-points and for indices $\alpha\neq\beta$ and
$\gamma\neq\lambda$:
\bd
\bra f_\alpha g_\beta h_\gamma k_\lambda\ket_M=
\delta_{\alpha\gamma}\delta_{\beta\lambda}
\bigbras \bra f h\ket_\star \bra gk\ket_\star
\plus \bra f\ket_\star\bra g\ket_\star \bra h\ket_\star \bra k\ket_\star
\minus \bra f h\ket_\star \bra g\ket_\star \bra k\ket_\star
\minus \bra gk\ket_\star \bra f\ket_\star\bra h\ket_\star \bigkets
\ed
\bd
{}~~~~~~~~~~~~~~~~~~~~~~~~~~~
+~\delta_{\alpha\lambda}\delta_{\beta\gamma}
\bigbras \bra f k\ket_\star \bra gh\ket_\star
\plus \bra f\ket_\star\bra g\ket_\star \bra h\ket_\star \bra k\ket_\star
\minus \bra f k\ket_\star \bra g\ket_\star \bra h\ket_\star
\minus \bra gh\ket_\star \bra f\ket_\star \bra k\ket_\star \bigkets
\ed
\bd
+~{\rm terms~with~less~than~two}~\delta{\rm 's} \room
\ed
Only the terms with two Kronecker $\delta$'s can contribute
to $\bra G^2\ket_M$, due to the specific properties of the replicon
fluctuations. We now obtain
\bd
\frac{1}{2}\bra G^2\ket_M=
\frac{1}{4}J^4\bigbras\left[1\minus\bra\sigma\ket_\star^2\right]^2\bigkets
\sum_{\alpha\beta}[\delta
Q_{\alpha\beta}\plus 2R_0\delta R_{\alpha\beta}\minus R_0^2\delta
q_{\alpha\beta}]^2
\ed
\bd
+\frac{1}{4 J^4(1\minus q)^4}\bigbras\left[\bra H^2\ket_\star\minus \bra
H\ket^2_\star\right]^2\bigkets
\sum_{\alpha\beta}\delta q^2_{\alpha\beta}
\ed
\bd
+\frac{1}{2(1\minus q)^2}\bigbras
\left[1\minus\bra\sigma\ket_\star^2\right]
\left[\bra
H^2\ket_\star\minus\bra H\ket^2_\star\right]\plus\left[\bra\sigma
H\ket_\star\minus\bra\sigma\ket_\star\bra H\ket_\star\right]^2\bigkets
\sum_{\alpha\beta}[\delta R_{\alpha\beta}\minus R_0\delta q_{\alpha\beta}]^2
\ed
\bd
-\frac{1}{2(1\minus q)^2}\bigbras \left[\bra\sigma H\ket_\star \minus
\bra\sigma\ket_\star\bra H\ket_\star\ket\right]^2\bigkets
\sum_{\alpha\beta}\delta q_{\alpha\beta}[\delta
Q_{\alpha\beta}\plus 2R_0\delta R_{\alpha\beta}\minus R_0^2
\delta q_{\alpha\beta}]
\ed
\bd
-\frac{J^2}{1\minus q}\bigbras \left[1\minus\bra\sigma\ket^2_\star
\right]
\left[\bra\sigma
H\ket_\star\minus\bra\sigma\ket_\star\bra H\ket_\star\right]\bigkets
\sum_{\alpha\beta}[\delta R_{\alpha\beta}\minus R_0\delta
q_{\alpha\beta}]
[\delta Q_{\alpha\beta}\plus
2R_0\delta R_{\alpha\beta}\minus R_0^2\delta q_{\alpha\beta}]
\ed
\bd
+\frac{1}{J^2(1\minus q)^3}\bigbras \left[\bra\sigma H\ket_\star\minus
\bra\sigma\ket_\star\bra
H\ket_\star\right]\left[\bra H^2\ket_\star\minus \bra
H\ket_\star^2\right]\bigkets
\sum_{\alpha\beta}\delta q_{\alpha\beta}[\delta R_{\alpha\beta}\minus
R_0\delta q_{\alpha\beta}]
\ed
The various combinations of matrix fluctuations can be somewhat
disentangled by introducing the transformation
\bd
\delta \bQ = \minus R_0^2\delta\bq\minus 2\frac{R_0}{J}\delta\br \plus
\frac{2}{J^2}\delta\bk
{}~~~~~~~~
\delta \bR= R_0\delta\bq\plus \frac{1}{J}\delta \br\room
\ed
In addition this renders all fluctuations dimensionless.
Expression (\ref{eq:repliconshift}) now acquires the
form
\be
\Psi-\Psi_{\rm RS}=\sum_{\alpha\beta}
\left(\!\!\begin{array}{c}
\delta k_{\alpha\beta}\\ \delta q_{\alpha\beta} \\ \delta
r_{\alpha\beta}
\end{array}\!\!\right)\bM
\left(\!\!\begin{array}{c}
\delta k_{\alpha\beta}\\ \delta q_{\alpha\beta} \\ \delta
r_{\alpha\beta}
\end{array}\!\!\right)
+\order(\delta^3)
\label{eq:finalrepliconshift}
\ee
in which the entries of the symmetric $3\times3$ matrix $\bM$  are
\be
\begin{array}{rl}
M_{11}= & \room
\bigbras\left[1\minus\bra\sigma\ket_\star^2\right]^2\bigkets \\
M_{12}=M_{21}=  & \room
\frac{1}{2}\minus\frac{1}{2J^2(1\minus q)^2}\bigbras\left[\bra\sigma
H\ket_\star\minus\bra\sigma\ket_\star\bra H\ket_\star\right]^2\bigkets
\\
M_{13}=M_{31}= & \room
-\frac{1}{J(1\minus q)}\bigbras
\left[1\minus\bra\sigma\ket^2_\star\right]\left[\bra\sigma
H\ket_\star\minus\bra\sigma\ket_\star\bra H\ket_\star\right]\bigkets
\\
M_{22}= & \room
\frac{1}{4(1\minus q)^2}\bigbras\left[1\minus\frac{\bra
H^2\ket_\star\minus\bra H\ket^2_\star}{J^2(1\minus
q)}\right]^2\bigkets-J^2R_0^2
\\
M_{23}=M_{32}= & \room
\frac{1}{2J^3(1\minus q)^3}\bigbras\left[\bra\sigma
H\ket_\star\minus\bra\sigma\ket_\star\bra H\ket_\star\right]\left[\bra
H^2\ket_\star\minus \bra H\ket^2_\star\right]\bigkets
-2J R_0 \\
M_{33}= & \room \frac{1}{2J^2(1\minus q)^2}\bigbras
\left[1\minus\bra\sigma\ket^2_\star\right]\left[\bra
H^2\ket_\star\minus \bra H\ket^2_\star\right]\plus\left[\bra\sigma
H\ket_\star\minus\bra\sigma\ket_\star\bra H\ket_\star\right]^2\bigkets
-1
\end{array}
\label{eq:matrixM}
\ee
We now use (\ref{eq:rsbR}) to eliminate the conjugate order
parameters $R_{\alpha\beta}$ from our equations. In the space of
RS saddle-points and replicon fluctuations we satisfy
$[\bR,\bq]=0$, so equation (\ref{eq:rsbR}) simplifies to
\bd
(\bq\bR)_{\alpha\beta}=\frac{i}{2J^2}\bra (H_\alpha\minus
J_0m\minus\theta)\sigma_\beta\ket_M
\ed
which after some algebra translates into the following
constraint on the replicon
fluctuations
\bd
M_{31}\delta\bk+M_{32}\delta\bq+M_{33}\delta\br=0
\ed
The stability of the RS saddle-point against replicon
fluctuations is now controlled by a symmetric $2\times 2$ matrix
$\overline{\bM}$:
\bd
\Psi-\Psi_{\rm RS}=\sum_{\alpha\beta}
\left(\!\!\begin{array}{c}
\delta k_{\alpha\beta}\\ \delta q_{\alpha\beta} \end{array}\!\!\right)
\overline{\bM}
\left(\!\!\begin{array}{c}
\delta k_{\alpha\beta}\\ \delta q_{\alpha\beta}
\end{array}\!\!\right)
+\order(\delta^3)
\ed
\bd
\overline{\bM}=
\left(\!\!\begin{array}{ccc}
1 & 0 & \minus M_{31}/M_{33}\room \\
0 & 1 & \minus M_{32}/M_{33}\room
\end{array}\!\!\right)
\bM
\left(\!\!\begin{array}{cc}
1 & 0 \room \\ 0 & 1 \room \\ \minus M_{31}/M_{33} & \minus M_{32}/M_{33}\room
\end{array}\!\!\right)
\ed
\be
=
\left(\!\!\begin{array}{cc}
M_{11}\minus M_{13}^2/M_{33} & M_{12}\minus M_{13}M_{32}/M_{33}\room \\
M_{12}\minus M_{13}M_{32}/M_{33} & M_{22}\minus M_{23}^2/M_{33}\room
\end{array}\!\!\right)
\label{eq:finalmatrixM}
\ee
Due to the curvature sign change of the second
derivative of $\Psi$, the analytic continuation to $n\rightarrow0$ of
the saddle-point that maximises $\Psi$ for $n>1$, will minimise $\Psi$
for $n<1$. This is emphasised explicitly by the summation over
$n(n\minus 1)$ non-trivial terms (all index combinations with
$\alpha\neq \beta$) in (\ref{eq:finalrepliconshift}). We can conclude
that the AT
instability occurs when the largest eigenvalue of the matrix
$\overline{\bM}$ is zero.

\subsection{Equilibrium}

{}From our previous result, the confirmation that the general (RSB)
thermodynamic
equilibrium state is a stationary state of our flow equation
(\ref{eq:finaldiffusion}), it follows that the same must hold within
the RS ansatz. We will now show this explicitly, as a non-trivial
consistency test (rather than a new result). The previous ansatz
(\ref{eq:equilansatz}) translates into
\be
\chi(\sigma,H)=e^{\frac{1}{2}\beta\sigma(H+\theta)}
\label{eq:RSequilansatz}
\ee
Due to (\ref{eq:RSequilansatz}) the measure $M_{\rm RS}$ in
(\ref{eq:rspsi}) becomes a Gaussian function of the fields, which
enables us to perform the field integrals and work out the RS
saddle-point equations (\ref{eq:RS2}-\ref{eq:RS7}). The result is:
\be
\mu=\frac{1}{2}\beta J_0m~~~~~~~~
Q=0~~~~~~~~
R_0=\frac{1}{2}\beta
\label{eq:RSequilsaddle1}
\ee
\be
m = \int \! Dz~ \tanh \beta(J_0 m\plus\theta\plus J z\sqrt{q})
\label{eq:RSequilsaddle2}
\ee
\be
q = \int \! Dz~ \tanh^2 \beta(J_0 m\plus\theta\plus J z\sqrt{q})
\label{eq:RSequilsaddle3}
\ee
which are the familiar RS equilibrium saddle-point equations, as
first obtained in
\cite{SK}.

We now turn to the right-hand side of equation
(\ref{eq:finaldiffusion}).
Since $Q=0$ the original two Gaussian variables $(x,y)$ in
(\ref{eq:rspsi}) are replaced by a single one, $z$.
With (\ref{eq:RSequilansatz}) and (\ref{eq:equilsaddle1}) we can
simplify the effective measure $M_{\rm RS}$ (\ref{eq:rsmeasure}), as
in
the full RSB case,
leading to
\bd
\bigbras \bra f[\sigma,H]\ket_\star\bra
g[\sigma,H]\ket_\star\bigkets\rightarrow
\int\!Dz ~\bra f[\sigma,H] \ket_\star\bra g[\sigma,H] \ket_\star
\ed
\bd
\bra f[\sigma,H] \ket_\star=
\frac{\sum_\sigma e^{\beta\sigma[J_0m+\theta+Jz\sqrt{q}]}
\int\!Dw~f[\sigma,J_0m\plus\theta\plus Jw\sqrt{1\minus q}\plus
Jz\sqrt{q}
\plus\beta J^2\sigma(1\minus q)]}
{2\cosh\beta[J_0m\plus\theta\plus Jz\sqrt{q}]}
\ed
In particular:
\be
\bra \delta[h\minus H]\delta_{\s,\sigma}\ket_\star=
\frac{e^{\beta\s h-\frac{1}{2}\beta^2 J^2(1\minus q)-\frac{1}{2 J^2(1-q)}
\left[h-J_0m-\theta-Jz\sqrt{q}\right]^2}}
{2J\sqrt{2\pi(1\minus q)}\cosh\beta[J_0m\plus\theta\plus Jz\sqrt{q}]}
\label{eq:staraveragedeltas}
\ee
The dependence of (\ref{eq:staraveragedeltas}) on $\s$ only through
a factor $e^{\beta\s h}$ immediately ensures that the first two terms
in the diffusion equation (\ref{eq:finaldiffusion}) cancel. Since this
happens even before we carry out the Gaussian average, we may write
\bd
\bra \delta[h\minus H]\delta_{\s,\sigma}\ket_\star=
\frac{1}{2}[1\plus \s \tanh(\beta h)]D(h;z)
\ed
implying relations like
\bd
\bra \sigma f(H)\ket_\star=\bra \tanh(\beta H)f(H)\ket_\star
\room
\ed
The building blocks of (\ref{eq:ooff}) thereby become
\bd
\bra\sigma\ket_\star=\tanh\beta[J_0m+\theta+Jz\sqrt{q}]
\room
\ed
\bd
\bra H\minus J_0m\minus\theta\ket_\star=
\beta J^2(1\minus q)\bra\sigma\ket_\star + Jz\sqrt{q}
\room
\ed
\bd
\int\!Dz \bra\tanh(\beta H)\ket_\star\bra\sigma\ket_\star=q
\room
\ed
\bd
\int\!Dz \bra\tanh(\beta H)\ket_\star\bra H\minus
J_0m\minus\theta\ket_\star
=2\beta J^2 q(1\minus q)
\room
\ed
\bd
\int\!Dz\bra\tanh(\beta H)(H\minus J_0m\minus\theta)\ket_\star=\beta
J^2(1\minus q^2)
\room
\ed
We will also need the following identity, obtained by
partial integration over $z$:
\be
\int\!Dz~z\bra\delta[h\minus H]\delta_{\s,\sigma}\ket_\star=
\frac{\sqrt{q}}{J}D(\s,h)(h\minus J_0m\minus \theta)-\beta
J\sqrt{q}(1\minus q)\int \! Dz\bra\delta[h\minus
H]\delta_{\s,\sigma}\ket_\star\bra\sigma\ket_\star\room
\label{eq:partialoverz}
\ee
We now have the necessary tools to analyse with minimum
effort the complicated
terms in our diffusion equation, given the ansatz
(\ref{eq:RSequilansatz}). The combined flow terms in
(\ref{eq:finaldiffusion}) can be simplified to
\bd
D(\s,h)(h\minus J_0m\minus\theta)D(\s,h)+
{\cal A}_{\rm RS}[\s,h;D]=
\left[1\minus\bra\tanh^2(\beta H)\ket_D\right]
\left[
\frac{1\minus 2q}{(1\minus q)^2}D(\s,h)(h\minus J_0m\minus \theta)
\right.
\ed
\bd
\left.
-\beta J^2\s D(\s,h)
+\frac{q\beta J^2}{1\minus q}\int\!Dz\bra\delta[h\minus
H]\delta_{\s,\sigma}\ket_\star
\bra\sigma\ket_\star
+\frac{Jq\sqrt{q}}{(1\minus q)^2}\int\!Dz~z\bra\delta[h\minus
H]\delta_{\s,\sigma}\ket_\star
\room
\right]
\ed
In order to evaluate the diffusion term in (\ref{eq:finaldiffusion})
we calculate the field derivative of $D(\s,h)$, using
(\ref{eq:staraveragedeltas}):
\bd
J^2\frac{\partial}{\partial h}D(\s,h)=
\beta J^2\s D(\s,h)-
\frac{h\minus J_0m\minus\theta}{1\minus
q}D(\s,h)+\frac{J\sqrt{q}}{1\minus q}\int\!Dz~z\bra\delta[h\minus
H]\delta_{\s,\sigma}\ket_\star
\ed
The full right-hand side of (\ref{eq:finaldiffusion}) can now be
written as
\bd
{\rm RHS}=
\left[1\minus\bra\tanh^2(\beta H)\ket_D\right]
\left[
\frac{q\beta J^2}{1\minus q}\int\!Dz\bra\delta[h\minus
H]\delta_{\s,\sigma}\ket_\star
\bra\sigma\ket_\star
-\frac{q}{(1\minus q)^2}D(\s,h)(h\minus J_0m\minus \theta)
\right.
\ed
\bd
\left.
{}~~~~~~~~~~~~~~~~~~~~~~~~~~~~~~~~~~~~~~~~
+\frac{J\sqrt{q}}{(1\minus q)^2}
\int\!Dz~z\bra\delta[h\minus H]\delta_{\s,\sigma}\ket_\star
\room
\right]=0
\ed
(by virtue of the identity (\ref{eq:partialoverz})).
The RS equilibrium state obtained in \cite{SK} therefore defines a
stationary state of our RS diffusion equation
(\ref{eq:finaldiffusion},\ref{eq:ooff}).
\vsp

Finally we turn to the AT instability, which we found to occur when
the largest eigenvalue of the matrix $\overline{\bM}$
(\ref{eq:finalmatrixM}) is zero. We can use the various
identities, derived for the thermal equilibrium state,
to simplify the matrix elements of $\overline{\bM}$
considerably:
\bd
\overline{\bM}_{\rm eq}=
\frac{\minus 1}{2(\beta J)^2\Lambda\minus 1}
\left(\!\!\begin{array}{cc}
\Lambda & \frac{1}{2}[1\minus (\beta J)^2\Lambda] \room \\
\frac{1}{2}[1\minus (\beta J)^2\Lambda] & \frac{1}{4}(\beta
J)^2[1\minus (\beta J)^2\Lambda]\room
\end{array}\!\!\right)
\ed
with
\bd
\Lambda=\int\!Dz~\cosh^{-4}\beta[J_0m\plus \theta\plus Jz\sqrt{q}]
\ed
The AT instability, as calculated within equilibrium statistical
mechanics \cite{AT}, occurs at $(\beta J)^2\Lambda=1$. Substitution of
this condition
into our expression for $\overline{\bM}_{\rm eq}$ immediately leads
to the desired result: the two eigenvalues of $\overline{\bM}_{\rm eq}$ are
$\{\minus\Lambda,0\}$, so the two conditions for the AT instability
coincide.

\section{Comparison with Simulations}

In order to verify the predictions of our theory we here compare the
results of solving numerically the (macroscopic) diffusion
equation (\ref{eq:finaldiffusion}), in which the disorder-generated
term ${\cal A}$ is calculated within the RS ansatz
(\ref{eq:ooff}), with the results of performing numerical simulations
of
the discretised version
of the underlying microscopic
stochastic dynamics (\ref{eq:master},\ref{eq:ratesandfields}).
Solving the diffusion equation (\ref{eq:finaldiffusion}), requires
making a discretisation not only of time, but also of the
joint spin-field distribution,
i.e. replace the two continuous functions $D_t(\pm 1,h)$  by two
histograms.  Furthermore, at each
time-step we have to solve the RS saddle-point equations
(\ref{eq:RS1}-\ref{eq:RS7}), which involve nested Gaussian
integrations.
It will be clear that the solution of equation
(\ref{eq:finaldiffusion})
requires a significant computational effort, even within the RS
ansatz, which is reflected in the
scope of the experiments described in this paper.
We restrict ourselves to describing the evolution
of the system in zero external field ($\theta=0$), following initial
states with individual spin states
chosen independently at random, given a required initial
magnetisation. Following the various experimental protocols that show
spin-glass ageing
phenomena, such as relaxation following cooling in a small field, and
relaxation with intermittant temperature increases or decreases, we
consider to be beyond the scope of this paper.

\subsection{Transients}

\begin{figure}[t]
\vspace*{85mm}
\hbox to
\hsize{\hspace*{-0cm}\includegraphics{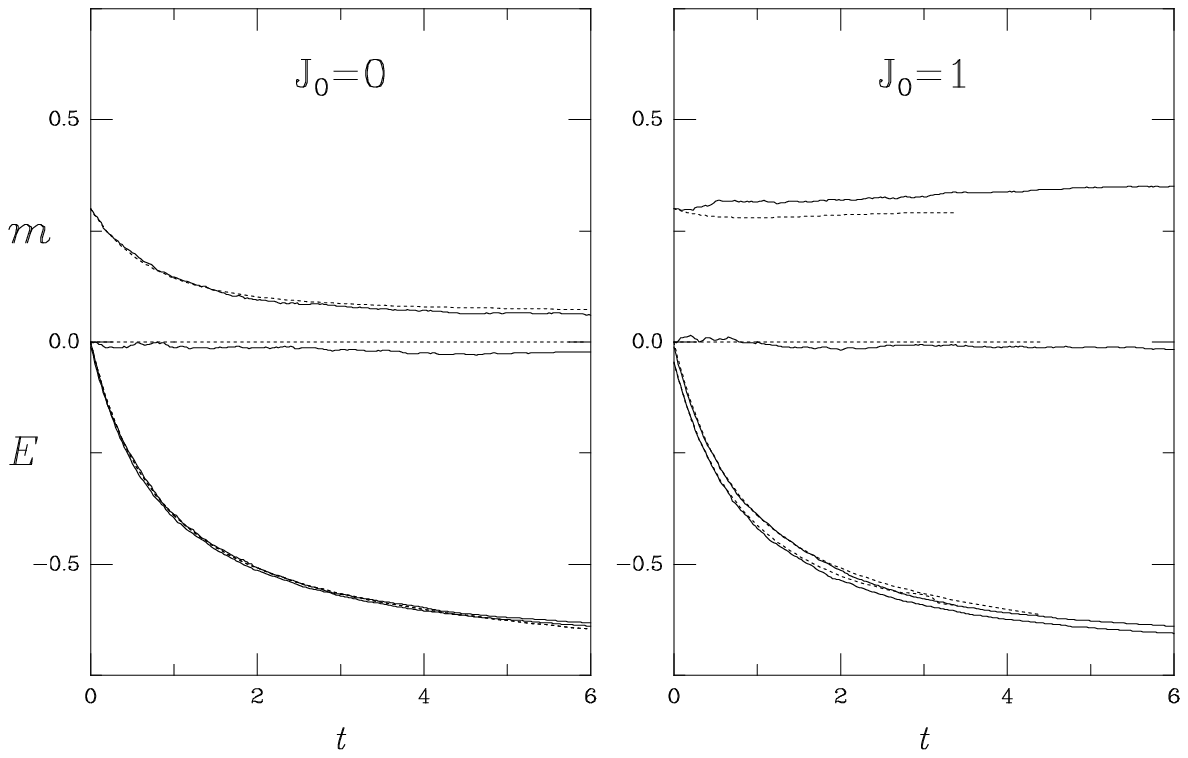}\hspace*{0cm}}
\vspace*{-10mm}
\caption{Evolution at $T=0$ of the magnetisation $m$ and the energy
per spin $E$, for $J_0=0$ (left) and $J_0=1$ (right) . Solid lines:
numerical simulations with $N=8000$; dotted lines: result of solving
the RS diffusion equation.}
\label{fig:flowT0}
\vspace*{105mm}
\hbox to
\hsize{\hspace*{-0cm}\includegraphics{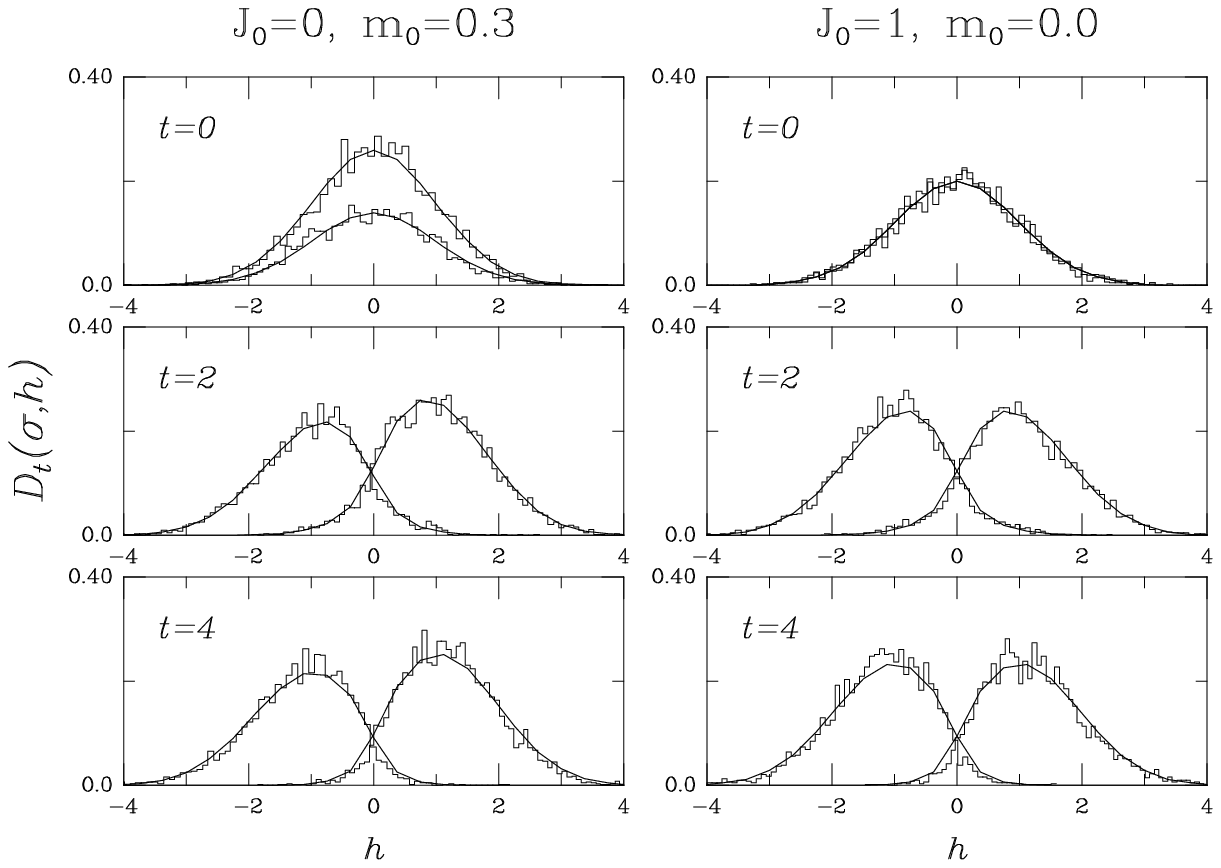}\hspace*{0cm}}
\vspace*{-10mm}
\caption{Evolution at $T=0$ of the two field distribitions
$D_t(\sigma,h)$,
for $J_0=0$ (left) and $J_0=1$ (right) . Histograms:
numerical simulations with $N=8000$; lines: result of solving
the RS diffusion equation. }
\label{fig:distT0}
\end{figure}
\begin{figure}[t]
\vspace*{85mm}
\hbox to
\hsize{\hspace*{-0cm}\includegraphics{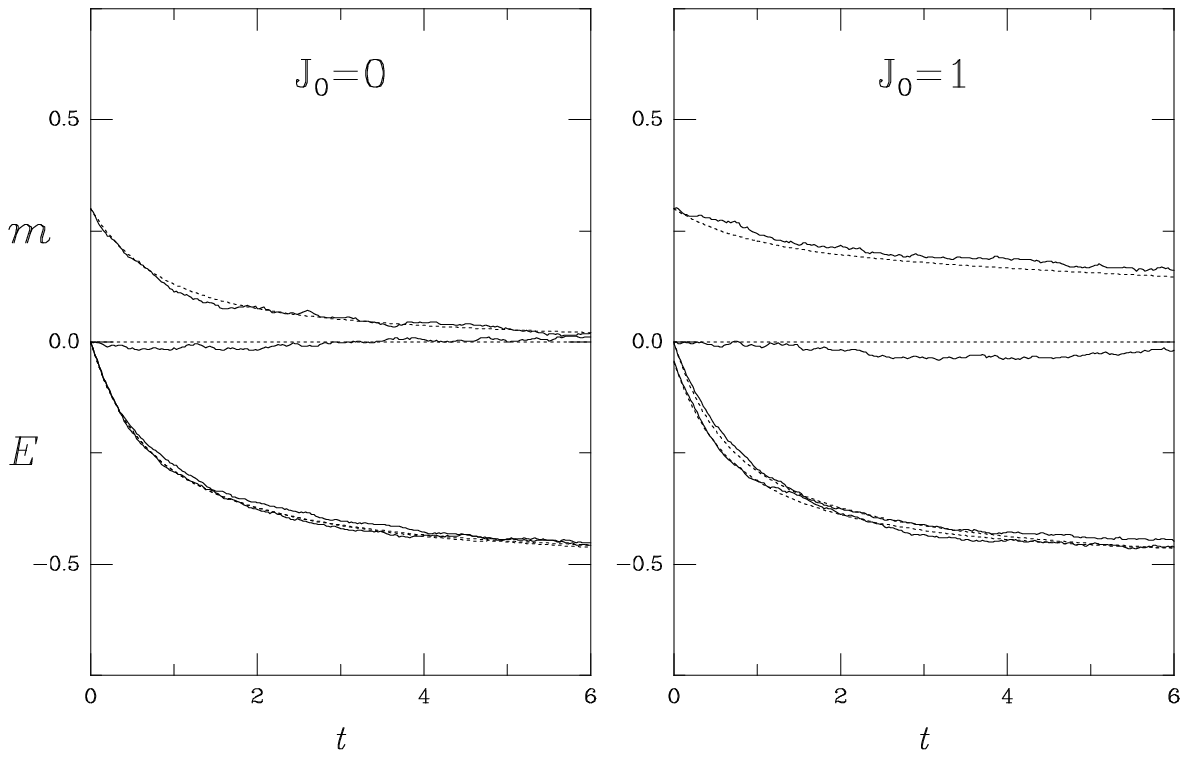}\hspace*{0cm}}
\vspace*{-10mm}
\caption{Evolution at $T=1$ of the magnetisation $m$ and the energy
per spin $E$, for $J_0=0$ (left) and $J_0=1$ (right) . Solid lines:
numerical simulations with $N=8000$; dotted lines: result of solving
the RS diffusion equation.}
\label{fig:flowT1}
\vspace*{105mm}
\hbox to
\hsize{\hspace*{-0cm}\includegraphics{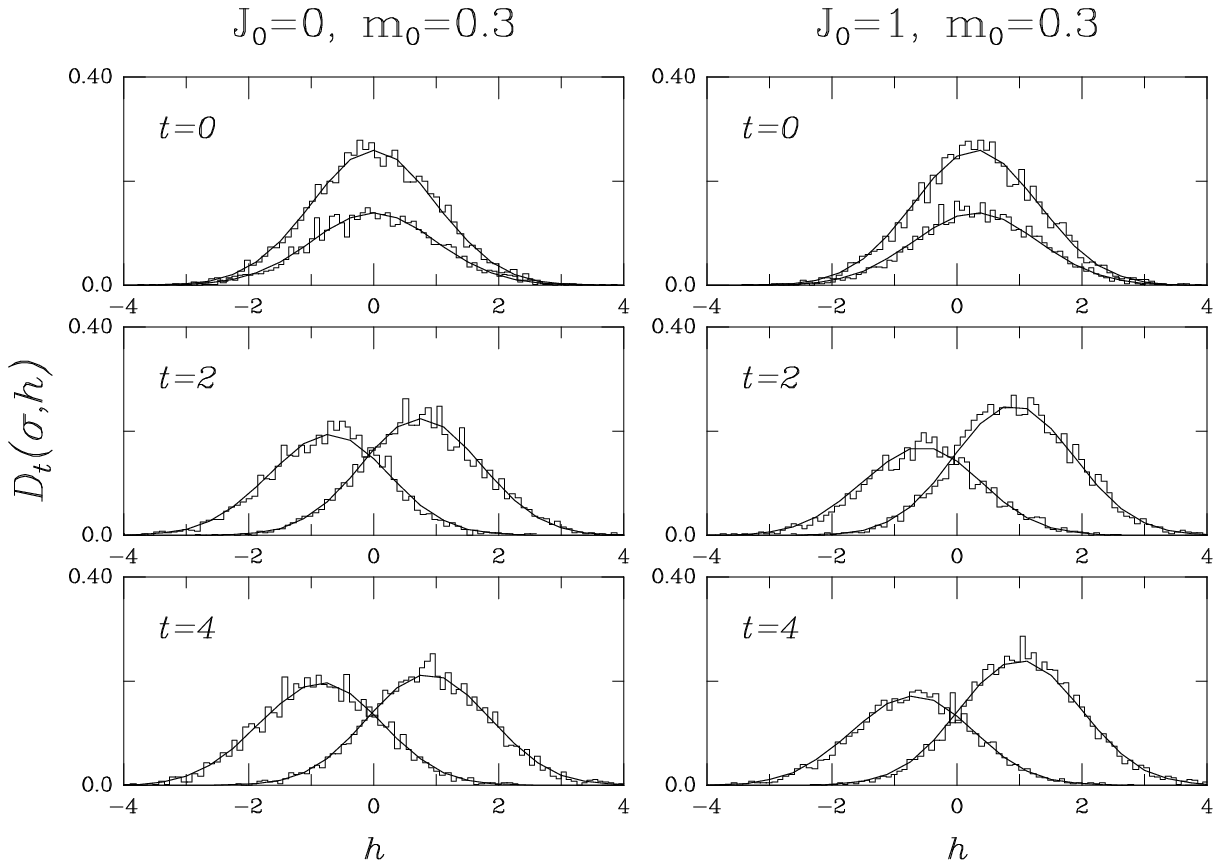}\hspace*{0cm}}
\vspace*{-10mm}
\caption{Evolution at $T=1$ of the two field distribitions
$D_t(\sigma,h)$,
for $J_0=0$ (left) and $J_0=1$ (right) . Histograms:
numerical simulations with $N=8000$; lines: result of solving
the RS diffusion equation. }
\label{fig:distT1}
\end{figure}

First we study the relaxation of the system on short time-scales. We
measure as a function of time the magnetisation $m$, the energy per
spin $E$, and the two distributions $D_t(\pm 1,h)$. Note that the full
local field distribution $D_t(h)$ is just the sum $D_t(1,h)\plus D_t(\minus
1,h)$. The numerical
simulations were
carried out with systems of $N=8000$ spins, following randomly drawn
initial states.
The results of confronting our theory with typical simulation
experiments,
for relaxations at $T=0$,
are shown in figures
\ref{fig:flowT0} and \ref{fig:distT0}, for $J_0=0$ (left pictures) and
$J_0=1$ (right pictures). In figure \ref{fig:flowT0} the top graphs
represent the magnetisation $m$ and the bottom graphs
represent the energy per spin $E$; for the two initial conditions
$m_0=0$ and $m_0=0.3$.
Figure \ref{fig:distT0} shows the corresponding distributions
$D(\sigma,h)$ for one particular choice of initial state ($D_t(1,h)$:
upper graph in
$t=0$ window, right graph in $t>0$ windows; $D_t(\minus 1,h)$: lower
graph in $t=0$ window, left graph in $t>0$ windows).
For $J_0=1$ we were not able to calculate the solution of equation
(\ref{eq:finaldiffusion}) up to $t=6$, due to the critical behaviour
of the saddle-point equations (\ref{eq:RS1}-\ref{eq:RS7}).
In figures \ref{fig:flowT1} and \ref{fig:distT1} we show similar
relaxation results for $T=1$.
As expected, at higher temperatures the two distributions $D_t(\pm 1,h)$
acquire a shape which becomes more like a Gaussian one, whereas in the
low temperature regime the deviations from a Gaussian shape become
important.

\begin{figure}[t]
\vspace*{75mm}
\hbox to
\hsize{\hspace*{10mm}\includegraphics{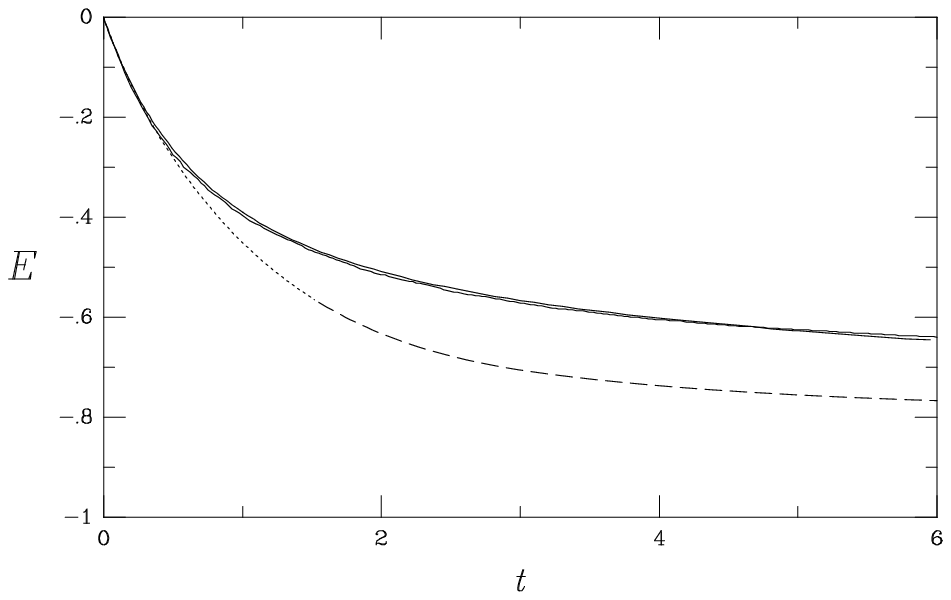}\hspace*{-10mm}}
\vspace*{-10mm}
\caption{Comparison of simulations (N=8000, solid line), the simple
two-parameter theory (RS stable: dotted line, RS unstable: dashed
line) and the advanced theory (solid line), for $T=J_0=0$. Note that
the two solid lines are almost on top of each other at the scale
shown.}
\label{fig:compare}
\end{figure}

To emphasise the increase in accurateness obtained by the present
advanced version
of our theory, as opposed to the simple two-parameter theory of
\cite{sk1}, we show in figure \ref{fig:compare} the simulation data
and
the predictions of the two
versions of our theory (simple versus advanced) corresponding to a
relaxation from a random initial state (with $m_0=0$), for
$T=J_0=0$. The failure of the two-parameter theory to account for
the typical slowing down of the dynamics
appears to have been amended convincingly by choosing as the dynamic
object the full distribution $D_t(\sigma,h)$, rather than just the
magnetisation and the energy per spin. Since the solution of our
diffusion equation (\ref{eq:finaldiffusion}), as depicted in figure
\ref{fig:compare}, is obtained within the RS ansatz, this slowing down
of the dynamics is not caused by replica symmetry breaking.
\clearpage

\subsection{Relaxation Near the SG Transition}

\begin{figure}[t]
\vspace*{86mm}
\hbox to
\hsize{\hspace*{10mm}\includegraphics{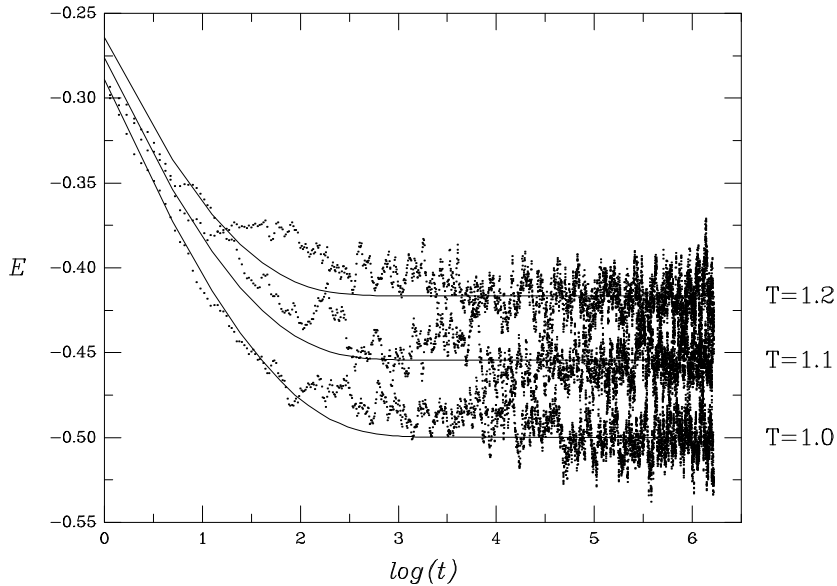}\hspace*{-10mm}}
\vspace*{-10mm}
\caption{Relaxation of the energy per spin $E$ for $J_0=\theta=0$
and $T\in\{1.0,1.1,1.2\}$. Dots:
numerical simulations with $N=3200$; lines: results of solving
the RS diffusion equation.}
\label{fig:diffusion}
\end{figure}
One way in which we can complement the short-time results presented so far,
whilst avoiding having to solve the saddle-point problem
(\ref{eq:RS1}-\ref{eq:RS7}) for large times, is to consider the
dynamics in the $q=0$ (paramagnetic) region. This allows us to
investigate the relaxation near $J_0=0$, $T=1$ (the critical
point which marks the $P\rightarrow SG$ transition).
In the
paramagnetic region the RS saddle-point problem can be solved,
\bd
q=m=R_1=0~~~~~~2J^2R_0=\bra\sigma H\ket_D~~~~~~J^4 Q^2=\bigbras\bra
H\ket^2_\star\bigkets
\ed
and the diffusion equation can be expressed entirely in terms of
(averages over) the distribution $D_t(\s,h)$ itself. Upon also
making use of the invariance of the problem with respect to an overall
spin sign change, we can write $D_t(\s,h)$ in terms of a single
function, the symmetric part of which is proportional to the local
field distribution:
\bd
D_t(\s,h)=\frac{1}{2}F_t(h)~~~~~~~~
\bra f(\sigma H)\ket_D=\int\!dy~F(y)f(y)=\bra f(y)\ket_F
\ed
In terms of $F_t$ the diffusion equation (\ref{eq:finaldiffusion}) becomes:
\bd
\frac{\partial}{\partial t}F_t(x)=
\frac{1}{2}[1\plus\tanh(\beta x)]F_t(\minus
x)-\frac{1}{2}[1\minus\tanh(\beta x)]F_t(x)
+J^2[1\minus\bra\tanh(\beta y)\ket_{F_t}]\frac{\partial^2}{\partial
x^2}F_t(x)
\ed
\be
+\frac{\partial}{\partial x}\left\{F_t(x)\left[\room
x[1\minus\bra\tanh(\beta y)\ket_{F_t}]\plus\bra y\ket_{F_t}\bra\tanh(\beta
y)\ket_{F_t}\minus\bra y\tanh(\beta y)\ket_{F_t}\right]\right\}
\label{eq:paradiffusion}
\ee
A randomly drawn initial state corresponds to
\bd
F_0(x)=\frac{1}{J\sqrt{2\pi}}e^{-\frac{1}{2}x^2/J^2}
\ed
Since (\ref{eq:paradiffusion}) is relatively easy to iterate
numerically, we can now compare the theoretical predictions with the
numerical data over much larger time-scales. In figure
\ref{fig:diffusion} we compare the result of solving
(\ref{eq:paradiffusion}) with numerical simulations, for
$t\in[0,500]$, in terms of the
energy per spin $E=\minus\frac{1}{2}\bra y\ket_{F_t}$.
We observe again a satisfactory agreement between theory and
experiment.

\section{Discussion}

The present paper is the second of a sequel in which we systematically
develop a  dynamical replica theory to describe the evolution of
macroscopic observables in the Sherrington-Kirkpatrick \cite{SK}
spin-glass. Our procedure for obtaining closed macroscopic flow
equations is based on two assumptions: $(i)$ the
flow equations are self-averaging with respect to the
realisation of the disorder, at any time, and $(ii)$ we may assume
equipartitioning
of probability in the macroscopic sub-shells of the ensemble.
The procedure can be shown to be {\em exact}, if the set of
macroscopic observables to which it is applied indeed obeys closed
dynamic equations. The resulting closed flow equations involve
a saddle-point problem, to be
solved at each instance of time, formulated in the replica language.

In our previous paper \cite{sk1} the closure procedure was
applied to the observables $m$ and
$E$ (the magnetisation and the energy per spin), resulting in a
two-parameter dynamical theory. Here we have shown how the same
procedure can be succesfully applied to the joint spin-field
distribution $D(\s,h)$, resulting in a dynamical theory describing an
infinite number of macroscopic order parameters.
The present, advanced, version of
our theory is again by construction exact for short times, in
equilibrium, and in the limit where the disorder is removed.
Furthermore, since the joint spin-field field distribution specifies
the
underlying
microscopic states in much more detail than would be the case by
specifying only the energy and the magnetisation (i.e. more microscopic
memory effects are being taken into acount), the equipartitioning
assumption has become much weaker.
We have restricted our analysis of the saddle-point
equations by making  the replica-symmetric (RS) ansatz.
On the time-scales considered in our simulation experiments,
the agreement between advanced RS theory and experiment is quite
satisfactory. For example, the slowing down missed by the
two-parameter theory is now well accounted for, and the theory
describes correctly the relaxation near  the spin-glass transition.
At this stage we  need more efficient numerical procedures in order
to extend the time-scales for which we can solve the equations of the
theory. This would enable us to compare, for instance, with data such
as the ones in \cite{kinzel}, to investigate the possible existence of
stationary states other than the one corresponding to thermal
equilibrium, and to see whether the theory can describe the typical ageing
phenomena observed in numerical simulations of similar mean-field
spin-glass models \cite{cugliandoloetal}.

The next and final stage of our program will be to investigate
for the Sherrington-Kirkpatrick spin-glass the effects of
replica symmetry breaking on the dynamic equations
\cite{sk3}. Although technically non-trivial, it is a straightforward
generalisation of the formalism developed so far.

Finally, a relevant question which we have not yet been able to answer
 is whether our diffusion equation
(\ref{eq:finaldiffusion},\ref{eq:rsbintensivepart}) is exact (for infinitely
large systems and on finite time-scales). There are several approaches
to this problem, each of which we plan to investigate in the near
future. The first approach is to apply our formalism to those
disordered spin systems for which the dynamics has been solved by
other means, like the non-symmetric SK model (in which each of the
bonds is drawn independently and asymmetrically at random
\cite{crisantisompolinsky,riegeretal}; preliminary results of this
study can be found in \cite{prl}), a toy model used in analysing
the shortcomings of the previous two-parameter approach
\cite{CF}, or the spherical spin-glass
\cite{cugliandolodean}. By definition, however, such exercises would
not yet prove exactness in the case of the SK spin-glass. The second
approach would be to try to derive a diffusion equation for the joint
spin-field distribution, starting from the equations for
correlation- and responsefunctions, as obtained from the path-integral
formalism
\cite{sommers}. The latter approach involves (rather
complicated) closed equations
for two functions $C(t,t^\prime)$ and $R(t,t^\prime)$, with two
real-valued arguments each (two times). The present
formalism also involves two functions $D_t(1,h)$ and $D_t(\minus 1,h)$, with
two real-valued arguments each (one time and one field).
It is therefore quite imaginable that both formalisms
constitute exact discriptions of the dynamics of the SK model.

\subsection*{Acknowledgement}

We would like to thank the Engineering and Physical Sciences Research
Council of the United Kingdom for partial support under grant GRH
73028 (ACCC) and scholarship 923.0910 (SNL).

\end{document}